\documentstyle [preprint,aps,eqsecnum,epsfig,fixes,floats]{revtex}

\makeatletter           
\@floatstrue
\def\figure{\let\@capwidth\columnwidth\@float{figure}}
\let\endfigure\end@float
\@namedef{figure*}{\let\@capwidth\textwidth\@dblfloat{figure}}
\@namedef{endfigure*}{\end@dblfloat}
\def\table{\let\@capwidth\columnwidth\@float{table}}
\let\endtable\end@float
\@namedef{table*}{\let\@capwidth\textwidth\@dblfloat{table}}
\@namedef{endtable*}{\end@dblfloat}

\makeatother

\def\Tc{T_{\rm c}}
\def\kB{k_{\rm B}}
\def\etal{{\it et al.}}
\def\eff{{\rm eff}}
\def\grad{\mbox{\boldmath$\nabla$}}
\def\x{{\bf x}}

\def\p{{\bf p}}
\def\k{{\bf k}}
\def\n{{\bf n}}
\def\U{{\rm U}}
\def\BZ{{\rm BZ}}
\def\lat{{\rm lat}}
\def\lunit{{\rm L}}
\def\phiphic{\langle\phi^2\rangle_{\rm c}}
\def\half{{\textstyle{1\over2}}}
\def\threehalf{{\textstyle{3\over2}}}
\def\fourth{{\textstyle{1\over4}}}
\def\c{{\rm c}}
\def\msb{{\overline{\rm MS}}}
\def\MSbar{$\msb$}
\def\phiz{\Phi}
\def\rz{r_0}
\def\uz{u_0}
\def\U{{\rm U}}
\def\I{{\rm I}}
\def\i{{\bf i}}

\def\eps{\epsilon}
\def\bare{{\rm bare}}
\def\gammaE{{\gamma_{\scriptscriptstyle{E}}}}
\def\cont{{\rm cont}}
\def\blparen{\mbox{\boldmath$($}}
\def\brparen{\mbox{\boldmath$)$}}

\def\reg{{\rm reg}}
\def\sing{{\rm sing}}
\def\Np{N_{\rm p}}

\def\BZ{{\rm BZ}}
\def\IL{I_{\rm L}}
\def\IC{I_{\rm C}}
\def\negsp{\hspace{-0.08in}}

\tightenlines
\begin {document}


\preprint {UW/PT-01-07}

\title {
   Monte Carlo simulation of O(2) \boldmath$\phi^4$ field theory
   in three dimensions
}

\author {Peter Arnold}

\address
    {%
    Department of Physics,
    University of Virginia,
    P.O. Box 400714,
    Charlottesville, VA 22904--4714
    }%

\author {Guy D. Moore}
\address
    {%
    Department of Physics,
    University of Washington,
    Seattle, Washington 98195--1560
    }%

\date {\today}
\maketitle

\begin {abstract}%
{%
Using standard numerical Monte Carlo lattice methods, we study
non-universal properties of the
phase transition of three-dimensional
$\phi^4$ theory of a 2-component
real field $\phi = (\phi_1,\phi_2)$ with O(2) symmetry.
Specifically,
we extract the renormalized values of $\langle\phi^2\rangle/u$ and
$r/u^2$ at the phase transition, where the continuum action of the theory is
$\int d^3x [ \half |\grad\phi|^2 + \half r \phi^2 + {u\over4!} \phi^4 ]$.
These values have applications to calculating the phase transition
temperature of dilute or weakly-interacting Bose gases
(both relativistic and non-relativistic).
In passing, we also provide perturbative calculations of various $O(a)$
lattice-spacing errors in three-dimensional O($N$) scalar field theory,
where $a$ is the lattice spacing.
}%
\end {abstract}


\section {Introduction}

A long standing problem is how to compute the first correction
$\Delta\Tc$, due to
interactions, to the critical temperature $\Tc$ for
Bose-Einstein condensation of a very dilute non-relativistic
homogeneous Bose gas in three dimensions.
In Ref.\ \cite{boselat1}, we review how this problem is related
to three-dimensional O(2) $\phi^4$ field theory \cite{baym1},
present the results
of lattice simulations of that theory, and so determine $\Delta\Tc$.
The purpose of the present work is to provide details of those simulations.
The lattice results presented here can also be applied to relativistic
Bose gases \cite{evans,tkachenko} and to non-relativistic gases in a trapping
potential \cite{trap}.

Three-dimensional O(2) $\phi^4$ theory has the continuum action
\begin {equation}
   S = \int d^3 x \left[ {1\over2}\, |\grad\phi|^2
         + {1\over2}\, r \phi^2 + {u\over 4!} (\phi^2)^2
       \right] ,
\label {eq:O2}
\end {equation}
where $\phi = (\phi_1,\phi_2)$ is a two-component real field
and $\phi^2 \equiv \phi_1^2 + \phi_2^2$.  For fixed $u$, we will
vary $r$ to reach the second-order critical point $r_\c(u)$ of this model.
The shift in the critical temperature of a non-relativistic homogeneous
single-species Bose gas is given in terms of this theory by \cite{baym1}
\begin {equation}
   {\Delta \Tc\over T_0} = - {2 m \kB T_0 \over 3\hbar^2 n} \,
        \Delta\phiphic ,
\label {eq:dTc}
\end {equation}
where $m$ is the boson mass, $n$ is the number density, and
\begin {equation}
   \Delta\phiphic \equiv \left[\phiphic\right]_u - \left[\phiphic\right]_0
\end {equation}
represents the difference between
the critical-point value of $\langle\phi^2\rangle$
for the cases of (i) $u$ non-zero and (ii) the ideal gas $u{=}0$ (with $r
\rightarrow 0$ from above).
The result for $\Delta\phiphic$ in the O(2) theory (\ref{eq:O2}) can only
depend on $u$ and so, by dimensional analysis, it must be proportional to
$u$.  A primary goal will be to discuss in detail the
measurement of the numerical constant $\Delta\phiphic/u$ from
lattice Monte Carlo simulations of the theory.  As reported in
Ref.\ \cite{boselat1}, our result is
\begin {equation}
   {\Delta\phiphic\over u} = -0.001198 \pm 0.000017.
\label {eq:finalphisqr}
\end {equation}

The critical value $r_\c$ of $r$ is also useful for various theoretical
applications, such as determining corrections to the value of the chemical
potential at the transition \cite{trap}, or to the critical temperature of a
trapped gas \cite{trap}, or to the critical temperature of
ultrarelativistic $\phi^4$ theory at zero chemical potential \cite{tkachenko}.
The $\phi^2$ interaction is associated with an ultraviolet (UV)
divergence of the three-dimensional theory and so must be renormalized.
If one chooses the renormalization scale $\bar\mu$ to be of order $u$ then,
by dimensional analysis, the renormalized value of $r_\c(\bar\mu)$ must be
proportional to $u^2$.  The precise scheme used to renormalize $r$, and
the precise choice of $\bar\mu$, is a matter of convention.  In this paper,
we will report a measurement of the numerical constant $r_\c/u^2$ for
$r$ defined by dimensional regularization and modified minimal subtraction
(\MSbar) at a renormalization scale of $\bar\mu=u/3$.  Our result is
\begin {equation}
   {r_\c^\msb(u/3) \over u^2} = 0.0019201 \pm 0.0000021 .
\label {eq:finalrc}
\end {equation}
One can convert to other choices of $\bar\mu$ by the (exact) identity
\begin {equation}
   {r^\msb(\bar\mu_1)\over u^2} = {r^\msb(\bar\mu_2)\over u^2}
      + {2\over9(4\pi)^2}\, \ln{\bar\mu_1\over\bar\mu_2} .
\label {eq:mubarconvert}
\end {equation}

In section \ref{sec:basics}, we discuss the lattice action we use,
its relationship to continuum fields and parameters, how we correct
for $O(a)$ lattice spacing errors, and the algorithms we
use for simulation.
Section \ref{sec:binder} details our procedure for finding the transition,
based on the method of Binder cumulants.
In section \ref{sec:finitesize}, we present our initial data, show that
we have simulated moderately large volumes,  and then
discuss how to analyze the remaining
finite-volume corrections by making use of the known critical
exponents associated with this universality class.
The corresponding numerical extrapolations of the finite-volume corrections
are given in section \ref{sec:Lunumerical}.
Numerical extrapolation of the continuum limit is presented in
section \ref{sec:uanumerical}.
A table of all our raw data for various size lattices and values of coupling
may be found in Appendix \ref{app:table}.
The derivations of the $O(a)$ lattice-spacing corrections used in this
paper are given in Appendix\ \ref{app:match}.
The remaining appendices include various discussions of scaling laws
used in the text, an analytic calculation of results for small lattice
volumes, and a critical discussion of one of the early simulations,
in the literature,
of the Bose-Einstein condensation temperature of dilute
non-relativistic gases.


\section {Lattice action, measurement, and algorithm}
\label {sec:basics}

The bare lattice Lagrangian has the form
\begin {equation}
   {\cal L} = a^3 \sum_\x \left[
      {1\over2} (- \phiz_1 \grad_\lat^2 \phiz_1 - \phiz_2 \grad_\lat^2 \phiz_2)
      + {\rz\over2} (\phiz_1^2+\phiz_2^2)
      + {\uz\over4!} (\phiz_1^2+\phiz_2^2)^2 
   \right] ,
\label{eq:S0}
\end {equation}
where $a$ is the lattice spacing.  In an actual simulation, one invariably
chooses lattice units where $a$=1, or equivalently works with
rescaled fields and parameters $\Phi^\lunit = a^{1/2} \Phi$,
$r_0^\lunit = a^2 r_0$, and $u_0^\lunit = a u_0$.  For the
sake of presentation, however, we will
generally avoid specializing to lattice units.

We work on a simple cubic lattice, and will work in cubic volumes
$L\times L\times L$ with periodic boundary conditions,
corresponding to $(L/a)^3$ sites.
The simplest possible implementation of the lattice Laplacian, which we call
the ``unimproved'' choice $\nabla_\U^2$, would be
\begin {equation}
\nabla^2_{\rm U} \Phi(\x) = a^{-2}
   \sum_\i \Bigl[\Phi(\x+a\i)-2\Phi(\x)+\Phi(\x-a\i)\Bigr] 
        , 
\label {eq:lapU}
\end {equation}
where the sum is over unit vectors in the three lattice
directions: (1,0,0), (0,1,0), and (0,0,1).
We use instead a standard improvement, which approaches the continuum
Laplacian faster for smooth fields:
\begin {equation}
   \nabla^2_{\rm I} \Phi(\x) = a^{-2} \sum_\i \Bigl[ {\textstyle
       -\frac{1}{12}\Phi(\x+2a\i)+\frac{4}{3}\Phi(\x+a\i)-\frac{5}{2}\Phi(\x)
       +\frac{4}{3}\Phi(\x-a\i)-\frac{1}{12}\Phi(\x-2a\i)
   } \Bigr] .
\label {eq:lapI}
\end {equation}
The difference can be seen from the Fourier transforms of the operators
$\nabla^2_\U$ and $\nabla^2_\I$, which are
\begin {eqnarray}
\label{eq:ktw_def}
      \tilde k_\U^2 &\equiv& a^{-2} \sum_i ( 2 - 2 \cos(a k_i) ) ,
\\
      \tilde k_\I ^2 &\equiv& a^{-2} \sum_i \left( \frac{5}{2} - \frac{8}{3} 
        \cos(a k_i) + \frac{1}{6} \cos(2 a k_i) \right) ,
\end {eqnarray}
and have small $k$ limits
\begin {eqnarray}
     \tilde k_\U^2 &=& k^2 + O(a^2 k^4)],
\\
     \tilde k_\I^2 &=& k^2 + O(a^4 k^6).
\end {eqnarray}
The unimproved Laplacian
has $O(a^2)$ error while the improved Laplacian has only $O(a^4)$ errors.


\subsection {An unimproved calculation of
             \boldmath$\Delta\langle\phi\rangle^2$}
\label {sec:div0}

One of our tasks is to calculate the continuum ratio
$\Delta\langle\phi^2\rangle/u$.
In the lattice theory (\ref{eq:S0}), the free field ($u_0=0$) result for
$\langle\phiz^2\rangle$ is
\begin {equation}
   \langle\phiz^2\rangle_{u_0=0}^{\phantom{1}} =
   \langle\phiz_1^2 + \phiz_2^2\rangle_{u_0=0}^{\phantom{1}} =
   2 \int_{\k\in\BZ} {1\over\tilde k^2} \,,
\end {equation}
where the integral is over the Brillouin zone $|k_i|\le\pi/a$.
Scaling out the dependence on $a$, we define
\begin {equation}
   {\Sigma\over 4\pi a} \equiv \int_{\k\in\BZ} {1\over\tilde k^2} .
\end {equation}
For the improved Laplacian, we obtain the value of the constant
$\Sigma$ by numerical integration:
\begin {equation}
   \Sigma \simeq 2.75238391120752 .
\end {equation}
On the lattice, the most straightforward implementation of the ratio
$\Delta\langle\phi^2\rangle/u$ is then
\begin {equation}
   {\Delta_0\langle\phiz^2\rangle \over u_0}
   \equiv
   {1\over u_0} \left[ \langle\phiz^2\rangle - {2\Sigma\over 4\pi a} \right] .
\label {eq:phi2renorm0}
\end {equation}
This will approach the desired continuum value as $ua\to0$, but the
lattice spacing errors at small $ua$ will be $O(ua)$.


\subsection {Relationship between lattice and continuum fields and parameters}
\label {sec:matchsummary}

One of our goals will be to reduce finite lattice spacing errors, so that
we can obtain better estimates of the continuum limit
with a given coarseness of lattice.
To eliminate errors at a given order in $a$, one must not only improve
the form of the Laplacian but must also perform an appropriate calculation
of the relationship between lattice and continuum parameters.
To this end, we will rewrite our bare lattice action in terms of
continuum fields $\phi$ and parameters ($r$,$u$) as
\begin {equation}
   {\cal L} = a^3 \sum_\x \left[
      {Z_\phi\over2}
           (- \phi_1 \grad_\lat^2 \phi_1 - \phi_2 \grad_\lat^2 \phi_2)
      + {Z_r (r+\delta r)\over2} (\phi_1^2+\phi_2^2)
      + {u+\delta u\over4!} (\phi_1^2+\phi_2^2)^2 
   \right] ,
\label {eq:Slat}
\end {equation}
where the renormalizations $Z_\phi$, $Z_r$, $\delta r$, and $\delta u$
depend on $r$ and $u$.  We explain
their derivation in Appendix \ref{app:match}.
For continuum $r$ defined by \MSbar\ renormalization at a scale $\bar\mu$,
we find
\begin {mathletters}
\label{eq:renorm1}
\begin {eqnarray}
   Z_\phi &=& 1 + \frac{2 C_2}{9} \, {(ua)^2 \over (4 \pi)^2}
       + O\blparen(ua)^3\brparen ,
\\
   Z_r &=& 1 + {2\xi\over 3} \, {ua\over (4\pi)}
       +  \left({4\xi^2\over9} - {2C_1\over3}\right) {(ua)^2\over(4\pi)^2}
       + O\blparen(ua)^3\brparen ,
\\
   \delta r &=& a^{-2} \left[
        - {2\Sigma \over 3}\, {(ua)\over(4\pi)}
        + {2 \over 9}
                \left( \ln{6\over \bar\mu a} + C_3 - 3 \Sigma\xi \right)
                {(ua)^2\over(4\pi)^2}
       + O\blparen(ua)^3\brparen 
     \right] ,
\label {eq:dr}
\\
   \delta u &=& a^{-1} \left[
        {5 \xi \over 3}\, {(ua)^2\over (4\pi)}
        + \left(-{32 C_1\over9} + \xi^2\right) {(ua)^3\over (4\pi)^2}
        + O\blparen(ua)^4\brparen 
      \right],
\label{eq:renormu}
\end {eqnarray}
\end {mathletters}%
where we have introduced several new numerical constants, given by various
integrals in lattice perturbation theory, whose values (for the action
with the improved Laplacian) are approximately
\begin {mathletters}%
\label {eq:consts}
\begin {eqnarray}
   \xi &\simeq& -0.083647053040968 ,
\\
   C_1 &\simeq& 0.0550612 ,
\\
   C_2 &\simeq& 0.0334416 ,
\\
   C_3 &\simeq& -0.86147916 .
\end {eqnarray}%
\end {mathletters}%
One needs to similarly match the operator $\phi^2$ whose expectation
is taken in determining $\Delta\langle\phi^2\rangle$.
In Appendix \ref{app:match},
we discuss the relationship between the continuum and lattice operators
and show that the continuum result for $\Delta\langle\phi^2\rangle$ is
\begin {equation}
\label {eq:renorm2}
   \Delta\langle\phi^2\rangle
   = Z_r \langle\phi^2\rangle_\lat - \delta\phi^2 + O(a^2),
\label{eq:phi2renorm}
\end {equation}
where the constant $\delta\phi^2$ is
\begin {equation}
   \delta\phi^2 = a^{-1} \left\{ {2\Sigma\over 4\pi} 
       + {4 \xi\Sigma \over 3}\, {ua\over(4\pi)^2}
       + {4\over9} \left[C_4-3\Sigma C_1 - \Sigma C_2 +2 \xi^2\Sigma
             + \xi \ln(\bar\mu a) \right] {(ua)^2\over (4\pi)^3}
       - 2 \xi \, {r a^2\over 4\pi} \right\} .
\label {eq:dphimaster}
\end {equation}
The new numerical constant is
\begin {equation}
   C_4 \simeq 0.282 .
\end {equation}

We note in passing that
the logarithm in (\ref{eq:dphimaster}) represents the explicit subtraction
of an effect analogous to a
[quadratic $\times$ log] correction
to the critical temperature for Bose condensation of dilute non-relativistic
gases, discussed in Ref. \cite{logs}.
In this analogy, the lattice spacing $a$ in our simulation plays
a role similar to the thermal wavelength in the Bose condensation problem,
and $u$ and $\bar\mu$ (which is chosen of order $u$) are proportional to
the scattering length.

In this paper, whenever we quote simulation results for a given value of
$ua$ and $r a^2$, we are referring to simulations of the action
(\ref{eq:Slat}) with parameters given by (\ref{eq:renorm1}) and
(\ref{eq:renorm2}) with $O(\cdots)$ set to zero.
When we quote values of $\Delta\phiphic$, we will be quoting continuum
values, as given by (\ref{eq:renorm2}).
This will be adequate to reduce the lattice spacing error to $O(a^2)$
on individual measurements of $\Delta\phiphic/u$ and to $O(a)$ on
individual measurements $r_\c$.

The precise definition of continuum $r$, and its relationship to bare
lattice $r_0$, are in principle unnecessary if 
one's interest is only to determine $\Delta\phiphic$
--- one could simply find the critical value of $r_0$ at any given lattice
spacing and not worry about its relation to continuum definitions.
However, as a practical matter, knowing the relationship facilitates
using results at one lattice spacing to make a good initial guess of the
critical value of $r_0$ at a new lattice spacing.  And, as discussed
earlier, the continuum critical value of $r$ is of interest in its
own right.

\begin {figure}
\vbox{
   \begin {center}
      \epsfig{file=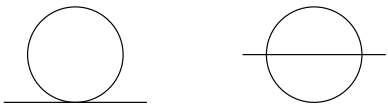,scale=.50}
   \end {center}
   \caption{
       The two fundamental UV-divergent graphs of continuum $\phi^4$ theory.
       \label{fig:3div}
   }
}
\end {figure}

Throughout this paper, continuum $r$ should be understood as
defined by \MSbar\ renormalization at a renormalization scale $\bar\mu$.
That is, the continuum Lagrangian is the $\eps\to 0$ limit of the
$(3{-}\eps)$-dimensional action
\begin {equation}
   S = \int d^{3-\eps}x \> \left[
             Z_\phi (\grad\phi)^*\cdot(\grad\phi)
             + r_\bare \phi^*\phi
             + \mu^\eps
                     {u_\eff\over 6} \, (\phi^*\phi)^2 \right] ,
\label {eq:O2ren}
\end {equation}
with the bare continuum parameter $r_\bare$ related to the renormalized
$r=r_\msb(\bar\mu)$ by
\begin {equation}
   r_\bare = r + {1\over (4\pi)^2\eps}\left(u\over3\right)^2 ,
\label{eq:rMSbar}
\end {equation}
and with
\begin {equation}
   \mu \equiv {e^{\gammaE/2}\over\sqrt{4\pi}} \, \bar\mu .
\label{eq:MSbar}
\end{equation}
[The factor of $e^{\gammaE/2}/\sqrt{4\pi}$
in (\ref{eq:MSbar}) is what distinguishes
modified minimal subtraction (\MSbar) from unmodified minimal subtraction
(MS); the difference between the two schemes amounts to nothing more
than a multiplicative redefinition of the renormalization scale.
The constant $\gammaE=0.5772\cdots$ is Euler's constant.]
Three-dimensional $\phi^4$ theory is super-renormalizable and the only
fundamental UV divergences of the continuum theory are those corresponding
to the two diagrams of
fig.\ \ref{fig:3div}.  The first has a linear divergence, which is ignored
by dimensional regularization.  The second has a logarithmic divergence,
which is the origin of the $u^2/\eps$ counter-term present in
(\ref{eq:rMSbar}).

\subsection {Algorithm}

We evolve configurations in Monte Carlo time using a combination of
site by site heatbath and multi-grid \cite{lattice} over-relaxation
updates. 

In section \ref{sec:binder}, we discuss how we define a nominal value
of the location $r_\c$ of the transition in finite volume.
In order to scan for the transition, we need to be able to smoothly
vary $r$.  We use the standard technique of canonical
reweighting.  Having made a simulation at one
value $r_{\rm sim}$ of $r$, and accumulated a 
Monte Carlo time series of values of
\begin {equation}
   \overline{\phi^2} \equiv {1\over V} \int d^3x\> \phi^2(\x) ,
\end {equation}
expectations at nearby values $r = r_{\rm sim} + \delta r$
can be calculated as
\begin {equation}
   \langle {\cal O} \rangle_r = \frac{
   \left\langle \exp\left(- \half V \, \delta r \, \overline{\phi^2}\right)
        {\cal O} \right\rangle_{r_{\rm sim}}}{
   \left\langle \exp\left(- \half V \, \delta r \, \overline{\phi^2}\right)
        \right\rangle_{r_{\rm sim}}} \, .
\label {eq:reweight}
\end {equation}

We estimate our statistical errors for the estimates of $r_\c$ and
$\Delta\phiphic$ in a given simulation run
using the jackknife method with 20 bins.
The size of bins must be large compared to the decorrelation time,
and this is verified in Appendix \ref{app:table}, where we quote
decorrelation times and sample sizes for each of our simulations.


\section {Finding the critical point}
\label {sec:binder}

Systems only have sharply defined phase
transitions in infinite volume, but we use the method of Binder cumulants
\cite{BinderCumulants}
to obtain a good estimate, in finite volume simulations, of the critical
value $r_\c$ of $r$.
Specifically, we measure the cumulant
\begin {equation}
   C = {\langle \bar\phi^4 \rangle \over \langle \bar\phi^2\rangle^2}
\end {equation}
as a function of $r$, where the 2-component vector
$\bar\phi$ is the
volume average of $\phi$:
\begin {equation}
   \bar\phi \equiv {1\over V} \int d^3x\> \phi(\x) .
\end {equation}
In infinite volume, the cumulant $C$ is 1 in the ordered phase and 2 in the
disordered phase.
In large volume, there is a smooth transition between these two values,
and the width of the transition region shrinks as the volume is
increased.
Specifically, for an $L\times L\times L$ volume, the width in $r$ of the
transition region scales as $L^{-y_t}$ in the $L \to \infty$ limit,
where $y_t = 1/\nu$ and $\nu$ is the correlation-length critical exponent.
The value of the exponent is
\begin {equation}
  y_t \equiv 1/\nu \approx 1.49
\end {equation}
for the universality class of the O(2) model.%
\footnote{
   \label{foot:alpha}
   The value of roughly 1.5 comes from the scaling relationship
   $\nu = (2-\alpha)/d$ and the fact that the specific heat critical
   exponent $\alpha$ is very small in this universality class.
   The best value of $\alpha$ is -0.01056(38) and
   comes from experiments in Earth orbit on superfluid
   He$^4$ \cite{alpha exp1,alpha exp2}.
   (See in particular endnote 15 of Ref.~\cite{alpha exp2}, and
   see also footnote 2 of Ref.~\cite{campostrini}.)
   For comparison, theoretical values (making use of $\alpha = 2-d\nu$
   as necessary) are $\alpha = -0.011(4)$, $-0.004(11)$, and $-0.0146(8)$
   from 3D series techniques \cite{guida},
   the $\epsilon$ expansion \cite{guida},
   and Monte Carlo \cite{campostrini}.
}

One method of estimating the location of the transition in finite volume
is to simply choose a fixed value $C_*$ of $C$ between 1 and 2 and then
define the nominal $r_\c$ in finite volume as the $r$ for which
$C(r) = C_*$.  This leads to errors in $r_\c$ (contributing
to errors in other quantities measured at the transition)
that would scale away as $L^{-y_t}$ in the large $L$ limit.
This $L^{-y_t}$ scaling of finite-size errors is typical of
many prescriptions for defining $r_\c$ in finite volume.

The method of Binder cumulants improves the scaling of the finite-size
error in $r_\c$ from $L^{-y_t}$ to $L^{-y_t-\omega}$, where
$\omega$ is the critical exponent associated with corrections to
scaling, and
\begin {equation}
   \omega \simeq 0.79
\end {equation}
for the universality class of the
O(2) model.%
\footnote{
   Values are $\omega=0.789(11)$, $0.802(18)$, and $0.79(2)$ from
   3D series techniques \cite{guida},
   the $\epsilon$ expansion \cite{guida},
   and Monte Carlo \cite{hasenbusch}.
}
One version of the method is to measure the curves $C(r)$ for
two different large system sizes $L_1$ and $L_2$, and then estimate
$r_\c$ as the point $r_\times$ where the curves intersect.
Specifically, Binder \cite{BinderCumulants} showed that, in the
$L_1, L_2 \to \infty$ limit, the error scales as
\begin {equation}
   r_\times(L_1,L_2) - r_\c
   \sim { L_2^{-\omega} - L_1^{-\omega} \over
          L_1^{y_t} - L_2^{y_t} }
   = { b^{-\omega} - 1 \over 1 - b^{y_t} } \, L_1^{-y_t-\omega} ,
\label{eq:rscaling}
\end {equation}
where $b \equiv L_2/L_1$.  Moreover, the value of the cumulant
$C_\times$ at the intersection approaches a {\it universal}\/ value $C_\c$ in
this limit.%
\footnote{
   For a nice numerical demonstration of this universality, see the
   Monte Carlo studies of Ising universality in Ref.\ \cite{blote}.
}

If one knew $C_\c$ in advance, then, for data in a given finite volume,
a nice method for determining a nominal point of transition is to
choose the $r$ such that $C(r) = C_\c$.  The finite size error in
$r_\c$ caused by this procedure also scales as $L^{-y_t-\omega}$.
This method is simpler and statistically a little cleaner than trying
to find the intersection of two $C(r)$ curves for two different values of $L$.
This is the method we shall use, but first we need a value of $C_\c$.
Because $C_\c$ is universal, its value can be measured from simulations of
any model in the same universality class.  We will use a value determined
by Campostrini \etal\ \cite{campostrini}, who
also simulate
a 2-component lattice $\phi^4$ theory.
They obtained $C_\c=1.2430(1)(5)$, where the two numbers in parenthesis
represent their statistical and systematic errors, respectively.
In the rest of this paper, when quoting results for the transition in
a given finite volume, we will mean the point where
\begin {equation}
    C(r) = 1.243
\label{eq:nominalCcrit}
\end {equation}
for that volume.
As a check, we have also made a
much less accurate determination of $C_\c$ using our own simulations,
which is discussed in Appendix \ref{app:Cc}.  We find $C_\c=1.2402(7)$,
which we suspect has a systematic bias, discussed in the Appendix.
The difference between these values of $C_\c$ is
insignificant for our purposes.  We have checked that the difference
would only change our final results by a tiny fraction of an error bar.


\section {Volume dependence of \boldmath$\Delta\phiphic$}
\label{sec:finitesize}

\begin {figure}
\vbox{
   \begin {center}
      \epsfig{file=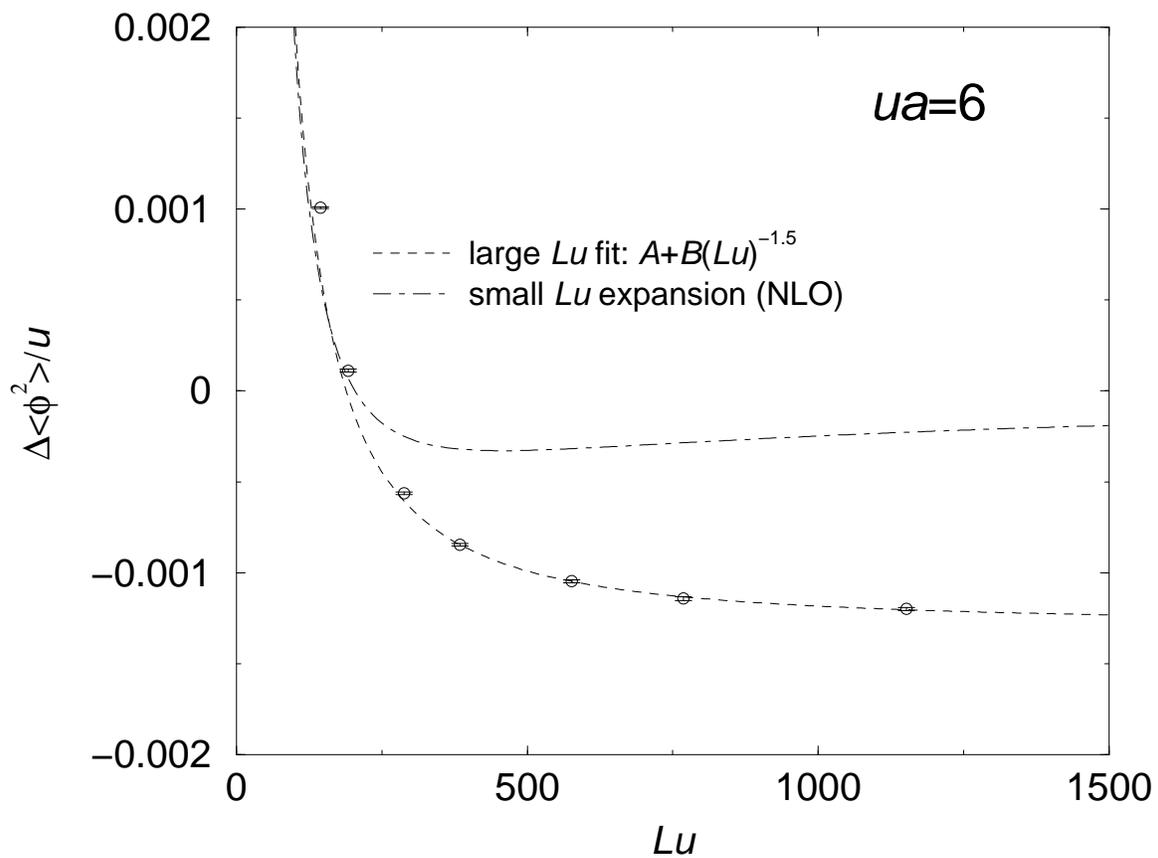,scale=.68,angle=-90}
   \end {center}
   \caption{
       Simulation results for $\Delta\phiphic/u$ vs.\ lattice size
       in physical units ($Lu$) for $ua=6$.  The large $Lu$ curve shown
       is a fit to the rightmost 4 data points, with confidence level
       61\%.
       \label{fig:ua6}
   }
}
\end {figure}

As discussed in the introduction, the only parameter of the continuum
problem at the critical point is $u$.  So the only length scale of the
problem is $1/u$.  The relevant measure of the size $L$ of a finite-volume
lattice relative to this scale is therefore $Lu$ (and similarly the relevant
measure of lattice spacing is $ua$).
Fig.\ \ref{fig:ua6} shows a plot of our results for
$\Delta\phiphic$ on $L\times L\times L$
lattices vs.\ $Lu$ for $ua=6$.
As can be seen, our largest $Lu$ values are clearly large:
the data clearly shows nice convergence towards an infinite volume limit.
To understand the size of the remaining finite-volume error, we will
want to fit the volume dependence at large $Lu$ to an appropriate
scaling law.

\subsection {Large volume scaling of \boldmath$\Delta\phiphic$}
\label{sec:Lscaling}

The scaling of large $L$ corrections depends on universal critical
exponents of the O(2) model, which is in the same universality class
as the classical $N=2$ Heisenberg ferromagnet.  The language of
critical exponents in the O(2) model is borrowed from correspondence
with the Heisenberg magnet, and $t \equiv r-r_\c$ is referred to as the reduced
``temperature'' in this context, $C \propto d^2(\ln Z)/dt^2$ as the
``specific heat,'' and so forth.
It is important to emphasize that this language holds the potential for
confusion because, in many applications (such as Bose-Einstein condensation
at fixed density), $t \equiv r-r_\c$ is not, in
fact, linearly related to the actual temperature of the system near
the critical point, and
${d^2(\ln Z)/dt^2}$ is not directly the physical specific heat.%
\footnote{
   A uniform,
   non-relativistic Bose gas
   is a constrained system: the particle density
   $n$ is fixed.  This constraint causes a different relationship of
   model parameters and the physical temperature than for unconstrained
   systems \cite{fisher}.  For example,
   the critical exponents
   $\tilde x = (\tilde\alpha, \tilde\beta, \tilde\gamma, \tilde\nu)$
   of the actual system are related to the standard
   exponents $x=(\alpha, \beta, \gamma, \nu)$ of the field theory by
   (i) $\tilde \alpha = -\alpha/(1-\alpha)$, and $\tilde x = x/(1-\alpha)$
   for the others, if $\alpha > 0$, or
   (ii) $\tilde x = x$ if $\alpha < 0$.
   This relation explains the difference between mean-field
   theory exponents for the O(2) model (e.g.\ $\alpha = 1/2$) and the
   exponents of a non-interacting Bose gas (e.g.\ $\tilde\alpha = -1$).
}
In any case, the analog of the ``energy density''
$E/V$ under this correspondence is
\begin {equation}
   {E\over V} \propto {d(\ln Z)\over V \, dt}
     = {1\over V} {d\over dr} \, \ln \int [{\cal D}\phi] e^{-S[\phi]}
     \propto \langle \phi^2 \rangle ,
\end {equation}
where $V = L^d$ is the system volume and $d=3$ is the dimension.
The problem of understanding how the finite volume corrections to
$\langle\phi^2\rangle$ scale with $L$ can therefore be stated in this
language as the problem of how corrections to the ``energy density'' scale with
$L$ for anything in this universality class.

The standard finite-size scaling ansatz provided by renormalization group
methods is that the free energy density $f = - V^{-1} \ln Z$ scales as%
\footnote{
   We have not included a ``magnetic field'' $h$ coupling to $\phi$,
   with a corresponding argument $b^{y_h} h$ in $f_\sing$, because we
   will only be interested in the case $h=0$ and are not interested in
   correlations of $\phi$ (as opposed to $\phi^2$), which could be generated
   by derivatives with respect to $h$.
}
\begin {equation}
   f(t,\{u_j\},L^{-1})
     = f_\reg(t,\{u_j\}) + b^{-d} f_\sing(b^{y_t} t, \{u_j b^{y_j}\}, b/L)
\end {equation}
for periodic boundary conditions,
where $f_\reg$ and $f_\sing$ generate the so-called regular and singular parts
of the free energy in the infinite volume limit.
(See Refs.\ \cite{barber,privman&fisher} and references therein.)
The length $b$ is an arbitrary renormalization scale (block size),
and $\{u_j\}$ denotes the set of infrared-irrelevant operators
(with corresponding $y_j < 0$).
Standard notation for critical exponents is $\nu=1/y_t$, and we will
denote the smallest $|y_j|$ associated with the irrelevant operators
$\{u_j\}$ as $\omega$.

Choosing $b = L$,
\begin {equation}
   f(t,{u_j},L^{-1})
     = f_\reg(t,{u_j}) + L^{-d} f_\sing(L^{y_t} t, {u_j L^{y_j}}, 1) .
\label {eq:f}
\end {equation}
The usual infinite-volume scaling form is obtained by taking
$L \to \infty$ with $t$ fixed, which, for the limit to exist, requires
\begin {equation}
   f_\sing(\tau,0,1) \sim \tau^{d/y_t} \qquad \mbox{as}
          \qquad \tau\to\infty .
\end {equation}
In contrast, for fixed $L$, the free energy will be analytic in $t$,
since there are no phase transitions in finite volume.
We can make a Taylor expansion of the finite-volume
free energy (\ref{eq:f}) in $t$ (as well as $\{u_j L^{y_j}\}$), which
should be a useful expansion when the arguments of the scaling piece
$f_\sing$ are small.  That is, for situations where $L^{y_t} t \to 0$ as
we take $L \to \infty$, we can Taylor expand (\ref{eq:f}) as
\begin {eqnarray}
   f &=& (A_0 + B_0 L^{-d} + C_0 L^{-d-\omega}) + \cdots)
\nonumber\\ &&
       + t (A_1 + B_1 L^{-d+y_t} + C_1 L^{-d+y_t-\omega} + \cdots)
\nonumber\\ &&
       + t^2 (A_2 + B_{20} L^{-d+2y_t} + \cdots)
\nonumber\\ &&
       + \cdots ,
\end {eqnarray}
where we have only displayed the leading corrections due to irrelevant
operators.  Differentiating with respect to $t$ to get the energy density,
we find that $\langle\phi^2\rangle$ scales in large volume as
\begin {eqnarray}
   \langle\phi^2\rangle &=&
       (A_1 + B_1 L^{-d+y_t} + C_1 L^{-d+y_t-\omega} + \cdots)
\nonumber\\ &&
       + 2 t (A_2 + B_{20} L^{-d+2y_t} + \cdots) 
       + \cdots .
\label{eq:e}
\end {eqnarray}

As mentioned earlier, use of the method of Binder cumulants to determine
$r_\c$ means that, in our application,
\begin {equation}
   t \sim L^{-y_t-\omega} .
\end {equation}
This indeed satisfies the condition $L^{y_t} t \to 0$ as $L \to \infty$,
and so the expansion (\ref{eq:e}) is appropriate.  For our application,
we then have
\begin {equation}
   \langle\phi^2\rangle =
       A_1 + B_1 L^{-d+y_t} + A_2' L^{-y_t-\omega}
           + C_1' L^{-d+y_t-\omega} + \cdots .
\end {equation}
Using the standard scaling relation $\alpha = 2-\nu d$ for the specific
heat scaling exponent $\alpha$, this can be rewritten in the form
\begin {equation}
   \langle\phi^2\rangle =
       A_1 + B_1 L^{-(1-\alpha)y_t} + A_2' L^{-y_t-\omega}
           + C_1' L^{-(1-\alpha)y_t-\omega} + \cdots .
\label{eq:ebinder}
\end {equation}
The value of $\alpha$ in the three-dimensional O(2) model is very small:
$\alpha \simeq -0.013$, corresponding to the value $y_t \simeq 1.49$
quoted earlier \cite{alpha exp1,alpha exp2,guida,hasenbusch}.

The renormalizations $Z_r$ and $\delta\phi^2$ that convert
$\langle\phi^2\rangle_\lat$ into $\Delta\langle\phi^2\rangle$
in (\ref{eq:phi2renorm})
do not introduce any new powers of $L$, and so the form of the
large-$L$ expansion of $\Delta\langle\phi^2\rangle$ is the same
as that for $\langle\phi^2\rangle$ in (\ref{eq:ebinder}) above,
though the coefficients are different.


\subsection {Large volume scaling for \boldmath $\alpha=0$}

Because we do not have large volume data spanning many decades in $L$,
$\alpha$ is zero for all practical purposes.  And so one might as well
use the $\alpha=0$ limit of the large-volume scaling (\ref{eq:ebinder}),
which corresponds to $y_t = d/2$.  Typically, $\alpha=0$ generates
logarithms in an RG analysis, which can appear as a superposition
\begin {equation}
   \lim_{\alpha\to0} {s^{z+q\alpha} - s^z \over \alpha} = q s^z \ln s
\end {equation}
of power laws $s^{z+q\alpha}$ and $s^z$ for some variable of interest $s$.
One might therefore expect that the $\alpha \to 0$ limit of the
large-volume scaling (\ref{eq:ebinder}) is
\begin {equation}
   \langle\phi^2\rangle =
       A_1 + B_1 L^{-d/2} + L^{-d/2-\omega} (C \ln L + D)
       + \cdots .
\label{eq:ebinder0}
\end {equation}
It's useful to verify the presence of a logarithm.  If it weren't
there, we could include the first corrections to scaling in large $L$ fits
of our data using a 3-parameter fit [$A_1$, $A_2$, $D$] rather than a
4-parameter fit [$A_1$, $B_1$, $C$, $D$; or equivalently $A_1$, $B_1$,
$A_2'$, $C_1'$ in (\ref{eq:ebinder})].

The existence of the logarithm can be directly related to the well known
logarithmic divergence of specific heat with $t$ when $\alpha = 0$.
We give a renormalization group analysis in Appendix \ref{app:log} that
makes this explicit.  Here, let us just note that the logarithm follows
from an old proposal by
Privman and Rudnick \cite{privman&rudnick}.
Ignore corrections to scaling for the moment, and suppose that
at $\alpha = 0$
the free energy had the general form (\ref{eq:f}) discussed earlier:
\begin {equation}
   f(t,L^{-1}) \sim f_\reg(t) + L^{-d} f_\sing(t L^{d/2}) .
\label{eq:fnaive}
\end {equation}
In order to get a logarithmic divergence $\ln(t^{-1})$ of the specific heat
in the infinite volume limit, we need a term $t^2 \ln(t^{-1})$ in the free
energy in that limit.  So we must have
\begin {equation}
   f_\sing(\tau) \sim A \tau^2 \ln(\tau^{-1})
        \qquad \mbox{as} \qquad \tau\to\infty .
\end {equation}
But this would give $f \sim f_\reg(t) + A t^2 \ln(t^{-1} L^{-d/2})$, which
doesn't have a good $L\to\infty$ limit.  The solution is to suppose that
the $\alpha=0$ version of the free energy is instead
\begin {equation}
   f(t,L^{-1}) \sim f_\reg(t) + A t^2 \ln(L^{d/2})
       + L^{-d} f_\sing(t L^{d/2}) ,
\label {eq:privman}
\end {equation}
which is what Privman and Rudnick proposed.
Note that the new term is analytic as $t \to 0$ for $L$ fixed, as it must be.
The new term gives a $t \ln L$ contribution to the energy density $E/V$, which,
for the $t \sim L^{-y_t-\omega}$ of interest to us, gives rise to the
logarithm term in (\ref{eq:ebinder0}).


\subsection {How large is large volume?}
\label{sec:Lbig}

\begin {figure}
\vbox{
   \begin {center}
      \epsfig{file=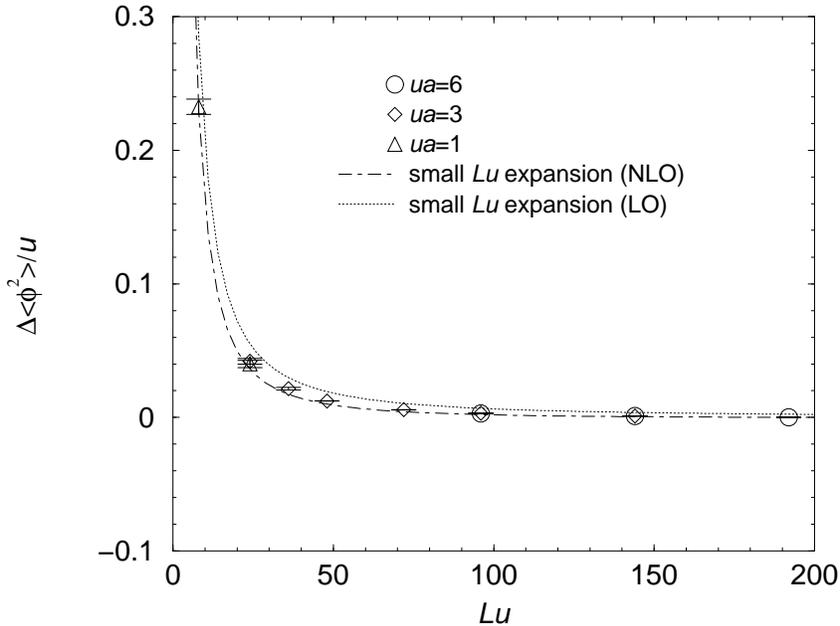,scale=.50,angle=-90}
   \end {center}
   \caption{
       Simulation results for $\Delta\phiphic/u$ vs.\ lattice size,
       showing more small volume data than Fig.\ \protect\ref{fig:ua6}.
       The dot-dash curve shows the next-to-leading order small $Lu$
       expansion of (\ref{eq:smallLu}).  The dotted curve shows just
       the leading-order term of that expansion.
       \label{fig:vsLu}
   }
}
\end {figure}

Before proceeding to numerical fits of the large volume dependence, it is
useful to first have an idea of how large $L$ should be before
one might reasonably hope for large volume scaling to hold.
{}From Fig.\ \ref{fig:ua6}, we see that the finite volume corrections
to the continuum value of $\Delta\phiphic$ become 100\% where the data
crosses zero, at roughly $Lu \sim 200$.  So one might guess that this is
very roughly the scale where scaling starts to set in---the scale that
separates short-distance perturbative physics from long-distance
critical-scaling
physics.  We can check this rough assessment from the other side.
In the limit of small $Lu$, the physics of fluctuations is perturbative,
and one can analytically compute the expected value of $\Delta\phiphic$
order by order in powers of $Lu$.  We perform this computation in
Appendix \ref{app:smallLu} in the continuum ($ua=0$) to next-to-leading
order (NLO) in $Lu$, with the result
that
\begin {equation}
   {\Delta\phiphic\over u}
   = {6.44003\over (Lu)^{3/2}} - {0.451570\over Lu}
      + O\blparen(Lu)^{-1/2}\brparen
\label {eq:smallLu}
\end {equation}
for our critical value (\ref{eq:nominalCcrit}) of the Binder cumulant.
The resulting curve is shown in both figs.\ \ref{fig:ua6} and
\ref{fig:vsLu} (as well as the leading-order small $Lu$ result in
fig.\ \ref{fig:vsLu}).  One can see that the small $Lu$ expansion becomes
unreliable past $Lu \sim 200$, which is the same as the previous
scale estimate.
(There is obviously nothing precise about this statement.
The scale at which the second term in the small $Lu$ expansion becomes
50\% of the first term, for example, is $Lu \sim 50$.)

\begin {figure}
\vbox{
   \begin {center}
      \epsfig{file=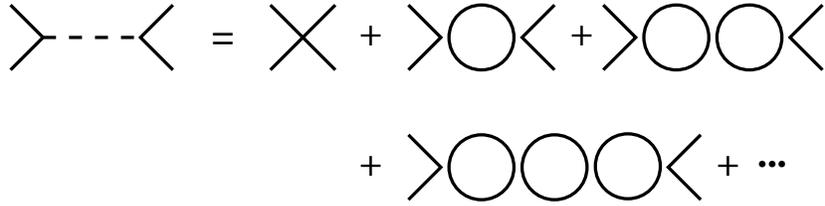,scale=.68}
   \end {center}
   \caption{
      Bubble chains, which are the source of scale dependence in
      large $N$ calculations of $\Delta\phiphic$ in O($N$) theory
      \protect\cite{largeN1,largeN2}.
      \label {fig:chain}
   }
}
\end {figure}

Interestingly, the scale $Lu \sim 200$ is close to
what one might estimate on the back of an envelope from a large-$N$
approximation to the theory.  In large $N$, one replaces the O(2) theory
studied here by an O($N$) theory of $N$ scalar fields and solves the theory
in the approximation that $N$ is large---a program pursued for the
problem of Bose-Einstein condensation of a non-relativistic gas in
Refs.\ \cite{largeN1,largeN2}.  In the $N \to \infty$ limit, one introduces
an auxiliary field $\sigma$ whose propagator represents a geometric sum
of bubble diagrams, such as shown in fig.\ \ref{fig:chain}.  The
corresponding resummed propagator is proportional to
\begin {equation}
   {1\over {3\over Nu} + \tilde\Sigma_0(p)} ,
\label {eq:sigma}
\end {equation}
where $\tilde\Sigma_0(p)$ represents the basic massless bubble integral
\begin {equation}
   \tilde\Sigma_0(p) \equiv {1\over 2} \int_{\l} {1\over l^2 |\l+\p|^2}
                    = {1\over 16p}.
\end {equation}
The propagator (\ref{eq:sigma}) changes from its large momentum behavior
($\to$ constant) to its small momentum behavior ($\propto p$) at a scale
given by $3/Nu \sim \Sigma_0(p)$.  This corresponds to
$p \sim Nu/48$ and so distance scales of order
$L \sim 2\pi/p \sim 96\pi/Nu$.  Setting $N=2$, we find
$Lu \sim 48\pi \sim 150$.  Of course, one would never hope that an
estimation this crude would be useful beyond, at best, the factor of 2 level.

Our discussion of system size has implications for the validity of an
early numerical study of the critical temperature for Bose-Einstein
condensation for non-relativistic gases \cite{gruter}.  We discuss this
in Appendix \ref{app:gruter}.


\section {Numerical extrapolation of
          \boldmath$L\mbox{\lowercase{$u$}}\to\infty$}
\label{sec:Lunumerical}

\subsection {Extrapolating \boldmath$\Delta\phiphic$ to
          \boldmath$Lu \to \infty$}

\begin {figure}
\vbox{
   \begin {center}
      \epsfig{file=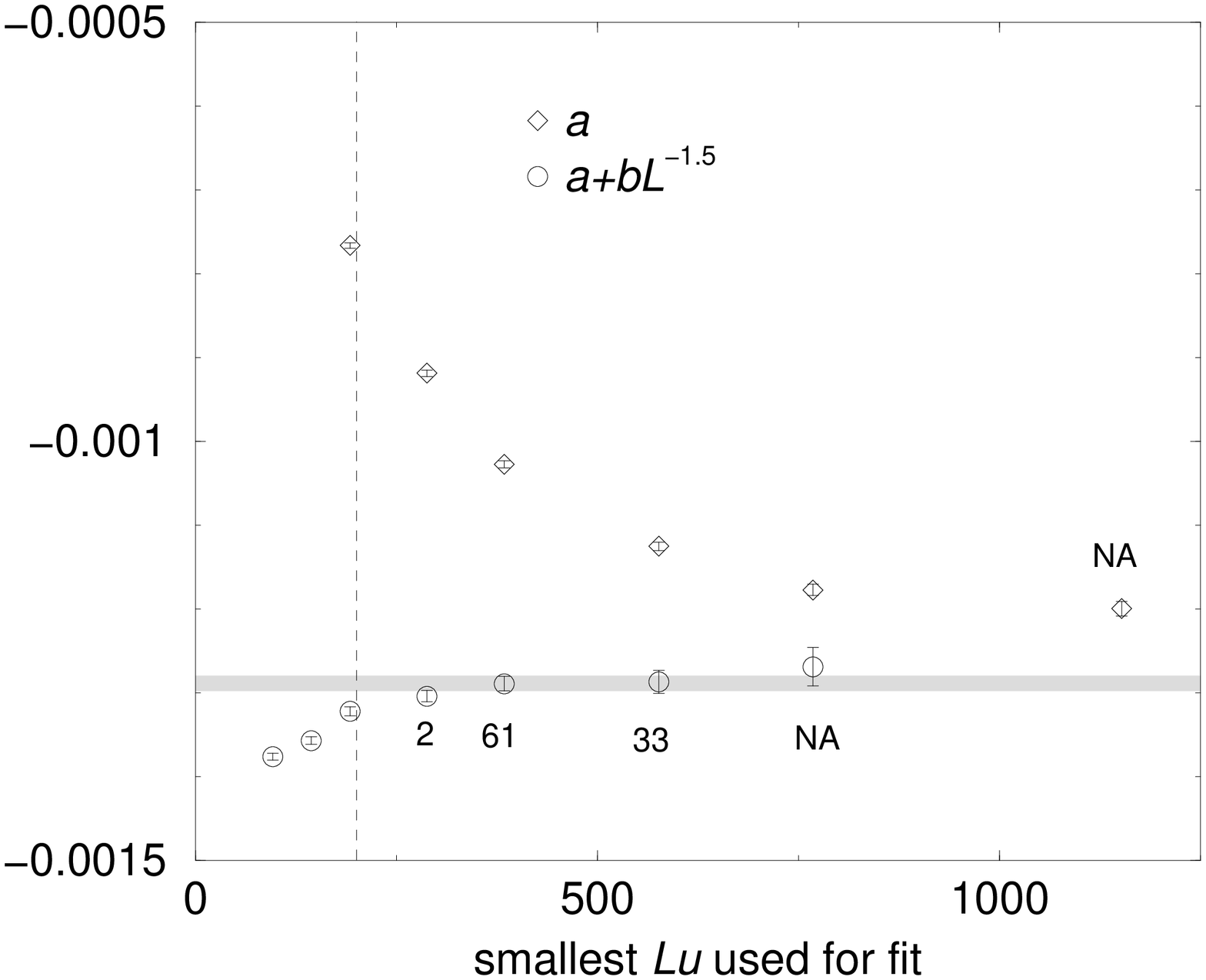,scale=.50,angle=-90}
   \end {center}
   \caption{
      Extrapolations of the $ua{=}6$ data of Fig.\ \ref{fig:ua6} to
      $L \to \infty$ under various assumptions of the functional form.
      The horizontal axis shows the smallest value of
      $Lu$ of the data used for the fit.
      Percentage confidence levels are written
      next to each extrapolation
      whenever non-negligible ($\ge$ 1\%),
      and ``NA'' (not applicable)
      refers to those cases where the number of data points
      used for the fit equals the number of parameters.
      \label {fig:ua6fits1}
   }
}
\end {figure}

We now examine fits of the $ua{=}6$ data of Fig.\ \ref{fig:ua6}
to the large $L$ scaling form of (\ref{eq:ebinder0}).
The circles in Fig. \ref{fig:ua6fits1} show the result of
extrapolating to
$L \to \infty$ using the leading-order scaling corrections
in (\ref{eq:ebinder0})---that is,
ignoring the sub-leading terms, which involve the exponent $\omega$.
For comparison, the
diamonds show what happens if we ignore finite-size effects altogether;
the corresponding confidence levels are terrible.
The vertical dashed line in the figure marks the
moderate system size $L = 200$ (discussed in the previous section),
below which you should become suspicious of any attempt to fit the
data to a large-volume scaling form.

A simple method for assigning a final result for the extrapolation
is to take our best fit with a correct scaling form.
As seen in Fig.\ \ref{fig:ua6fits1}, the leading-order scaling form
$A + B L^{-d/2}$ is perfectly adequate for fitting our large $Lu$ data.
Our procedure is to fit the data for lattice sizes greater than or equal to
a given $L_{\rm min}$, decreasing this minimum size for as long as the fit
remains stable with a reasonable confidence level.
We take as our estimate the 61\% confidence level fit to $Lu \ge 384$,
which gives
\begin {equation}
   \left[\Delta\phiphic\over u\right]_{ua=6} = -0.001289(9) ,
\label {eq:ua6limit}
\end {equation}
which is depicted by the shaded region of Fig.\ \ref{fig:ua6fits1}.

\begin {figure}
\vbox{
   \begin {center}
      \epsfig{file=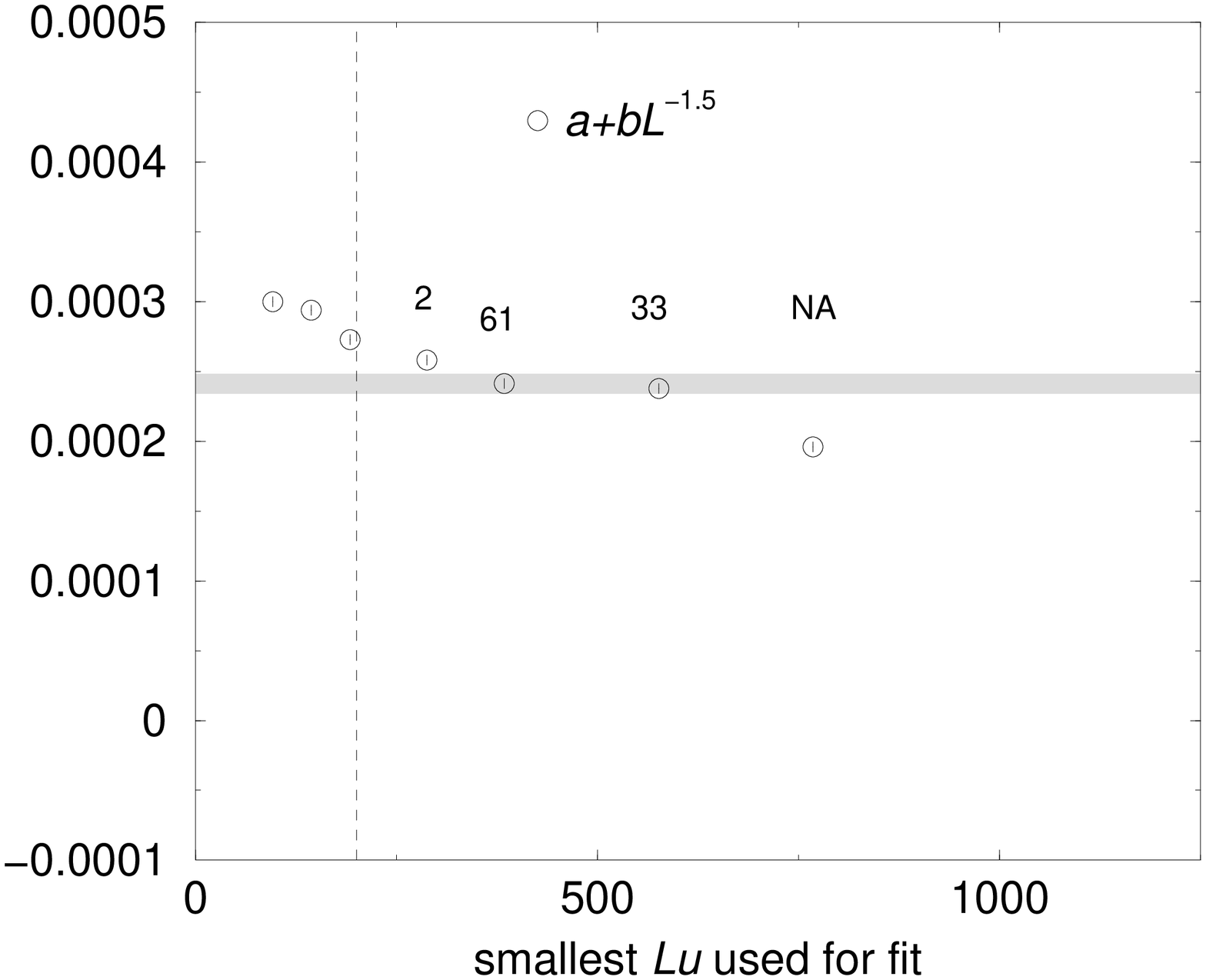,scale=.50,angle=-90}
   \end {center}
   \caption{
      As Fig.\ \ref{fig:ua6fits1} but shows the magnitude of the finite-size
      correction to $\Delta\phiphic/u$ at $Lu=576$, as determined by the
      fit.
      \label {fig:ua6diff576x1}
   }
}
\end {figure}

As described in Ref.\ \cite{boselat1}, we will actually use
$Lu=576$ as a reference point from which to derive finite volume
and finite lattice spacing corrections.
Fig.\ \ref{fig:ua6diff576x1} is
similar to Fig.\ \ref{fig:ua6fits1} but
shows the size of the finite-size correction at $Lu=576$, as
determined by the fit.  The best fit (the 61\% confidence level one)
gives
\begin {equation}
   \left[\Delta\phiphic \over u\right]_{Lu=576}
   - \left[\Delta\phiphic\over u\right]_{Lu\to\infty}
   = 0.000241(7),
\label{eq:Lcorrect576}
\end {equation}
and we will use this difference, rather than the limit
(\ref{eq:ua6limit}), in what follows.
[The difference is determined by the coefficient $B$ of our fit to
$A + B L^{-d/2}$, and so is determined by all the data points of that
fit; it is
not simply our $(Lu,ua)=(576,6)$ data point
minus the limit (\ref{eq:ua6limit}), which would produce a larger error.]
As we shall see, $ua=6$ is a reasonably small value of $ua$, and we expect
this to be a good estimate to the finite volume corrections in the
continuum ($ua\to0$) limit.

\begin {figure}
\vbox{
   \begin {center}
      \epsfig{file=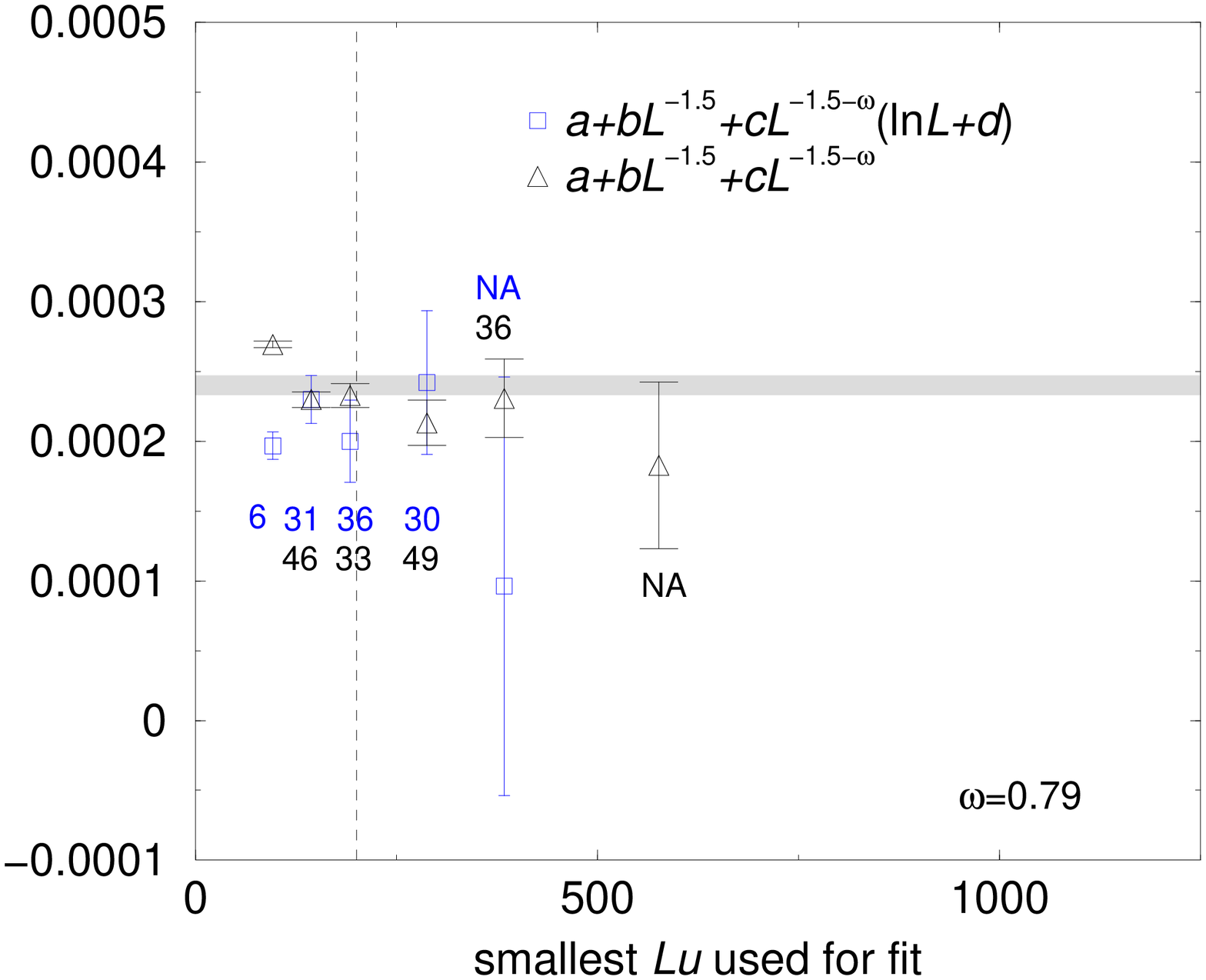,scale=.50,angle=-90}
   \end {center}
   \caption{
      As Fig.\ \ref{fig:ua6diff576x1} but showing extrapolations that include
      corrections to scaling.  Where two confidence levels are listed on
      top of each other, the upper one is for the square data point and the
      lower for the triangular data point.
      \label {fig:ua6diff576x2}
   }
}
\end {figure}

As a check that corrections to scaling will not radically alter our
results, we show as squares in Fig.\ \ref{fig:ua6diff576x2} the result of
fits to (\ref{eq:ebinder0}).  The values are consistent with the previous
result, with larger errors because we are fitting more parameters.
The triangles show a simplified fit, with one less parameter, that ignores
the logarithmic dependence.

\subsection {Extrapolating $r_\c$ to $Lu \to \infty$}

We will now make a similar analysis of finite-size effects for the
critical value $r_\c$ of $r$.  The continuum value of $r$ is convention
dependent---it depends on one's choice of renormalization scheme and
renormalization scale.  As discussed earlier, our convention will be
to define $r$ with \MSbar\ regularization at the renormalization scale
$\bar\mu=u/3$.
The conversion formula (\ref{eq:mubarconvert}) to other choices of
$\bar\mu$ can be extracted from (\ref{eq:dr}) and the fact that the
theory is super-renormalizable.

\begin {figure}
\vbox{
   \begin {center}
      \epsfig{file=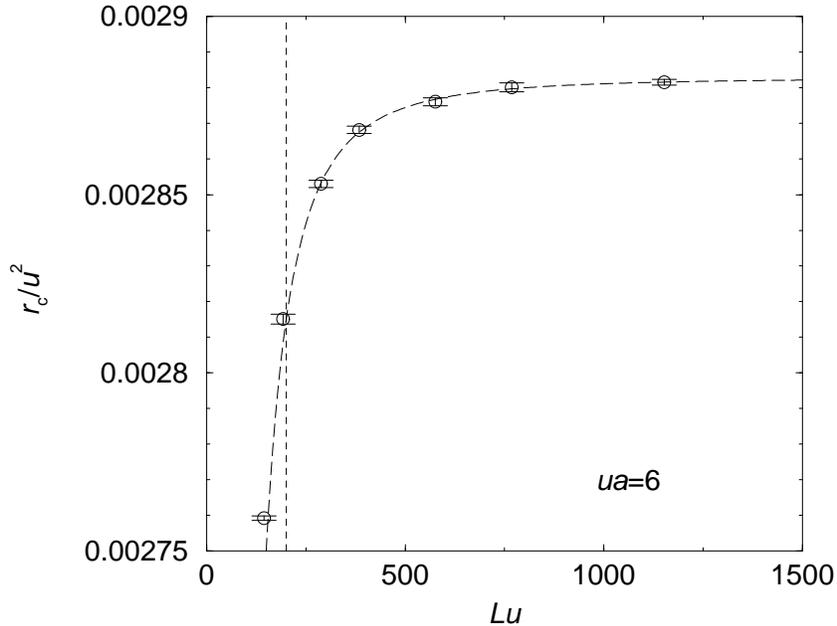,scale=.50,angle=-90}
   \end {center}
   \caption{
      $ua=6$ data for the nominal value of $r_\c/u^2$ (defined in
      \MSbar\ renormalization at
      renormalization scale $\bar\mu=u/3$) as a function of system size.
      The fit shown by the dashed line is the large-volume fit to all
      but the leftmost data point.
      \label {fig:ua6musqr}
   }
}
\end {figure}

\begin {figure}
\vbox{
   \begin {center}
      \epsfig{file=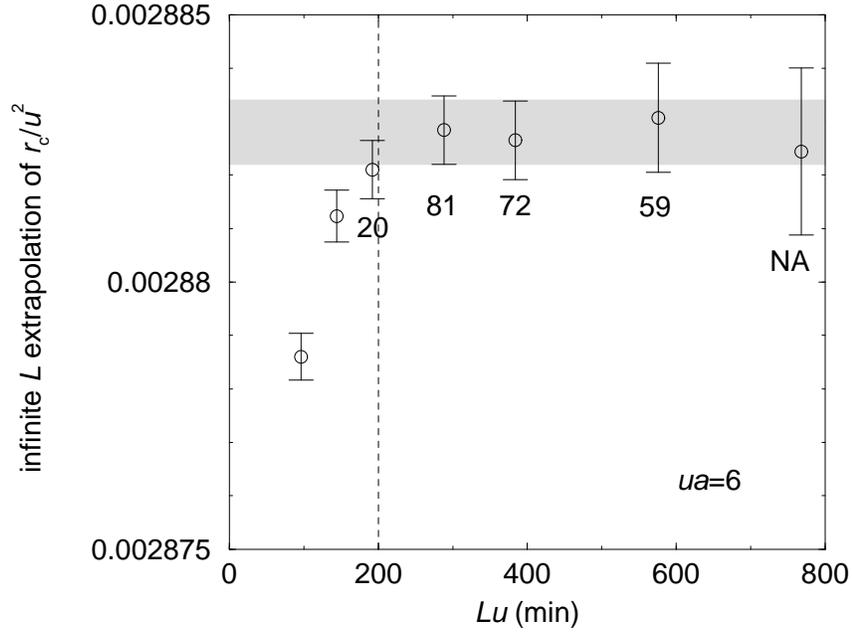,scale=.50,angle=-90}
   \end {center}
   \caption{
      Extrapolations of the $ua{=}6$ data of Fig.\ \ref{fig:ua6musqr} to
      $L \to \infty$ using the functional form $A + B L^{-y_t-\omega}$.
      The horizontal axis shows the smallest value of
      $Lu$ of the data used for the fit.
      Confidence levels are written as described for Fig.\ \ref{fig:ua6fits1}.
      \label {fig:ua6musqrfit}
   }
}
\end {figure}

\begin {figure}
\vbox{
   \begin {center}
      \epsfig{file=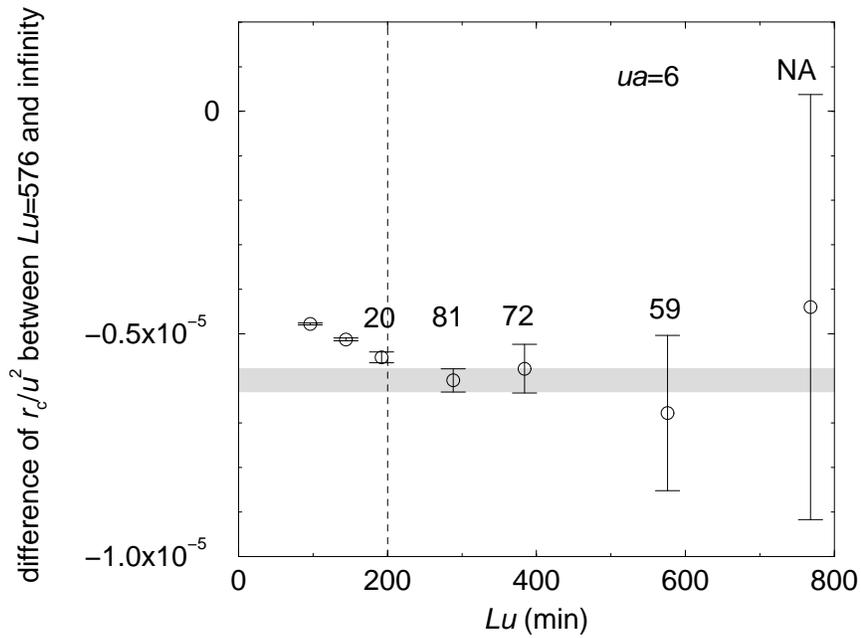,scale=.50,angle=-90}
   \end {center}
   \caption{
      As fig.\ \ref{fig:ua6musqrfit} but showing the
      magnitude of the finite-size correction to $r_c/u^2$ at $Lu=576$.
      \label {fig:ua6diff576musqr}
   }
}
\end {figure}

Fig.\ \ref{fig:ua6musqr} shows, for $ua=6$,
the dependence on system size of our
finite-volume determinations of $r_\c(u/3)/u^2$.  As discussed in section
\ref{sec:binder}, the finite-size corrections are expected to scale
as $L^{-y_t-\omega}$ as $L \to \infty$.
Fig.\ \ref{fig:ua6musqrfit} shows the result of extrapolating
an infinite-volume result by fitting various subsets of the
data to $A + B L^{-(d/2)-\omega}$.
Taking the highest confidence level fit,
\begin {equation}
   \left[r_\c(u/3)\over u^2\right]_{ua=6}
   = 0.0028828(6) .
\end {equation}
Fig.\ \ref{fig:ua6diff576musqr} shows the size of the finite volume
error at our canonical $Lu=576$, as determined by the fits,
\begin {equation}
   \left[r_\c(u/3)\over u^2\right]_{Lu=576}
   - \left[r_\c(u/3)\over u^2\right]_{Lu=\infty}
   = -0.00000604(26) .
\label {eq:Lcorrect576r}
\end {equation}

\section {Numerical extrapolation of \lowercase{\boldmath$ua \to 0$}}
\label {sec:uanumerical}

\subsection {Extrapolating $\Delta\phiphic$ to $ua \to 0$}

\begin {figure}
\vbox{
   \begin {center}
      \epsfig{file=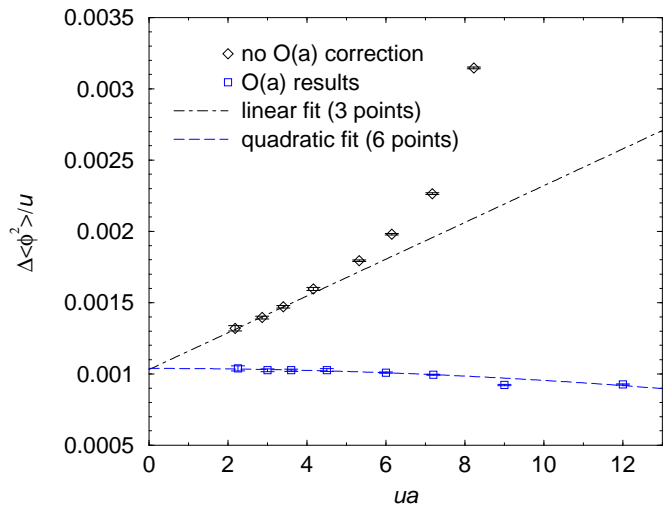,scale=.40,angle=-90}
   \end {center}
   \caption{
       The squares show results for $\Delta\phiphic/u$ vs.\ $ua$ at
       $Lu=144$.
       The line through them is a fit of the first 6 points to
       $A + B(ua)^2$.
       The diamonds represent the corresponding uncorrected data, as
       described in the text.  To guide the eye, a straight line has been
       fit through the first three points, though the actual small $ua$
       behavior is linear$\times$log.
       \label{fig:Lu144}
   }
}
\end {figure}

We will now discuss numerical results for the continuum limit $a \to 0$
while holding the physical volume of the lattice fixed.  This can
be expressed as $ua \to 0$ holding $Lu$ fixed.
Since the number $(L/a)^3$ of lattice sites we can practically include
in a simulation is limited,
we can obviously
explore smaller values of $ua$ when we fix smaller values of $Lu$.
As a test that we understand our lattice spacing errors, we have
therefore made several simulations at the rather moderate system size
of $Lu=144$ (see the discussion of sec.\ \ref{sec:Lbig}).
The results for the dependence of $\Delta\phiphic/u$
on $ua$ are shown by the
squares in Fig.\ \ref{fig:Lu144}.
If our corrections for $O(a)$ errors have been calculated correctly, the
remaining error should be $O(a^2)$.  Indeed, all but the largest two
$ua$ data points in Fig.\ \ref{fig:Lu144} fit very well the
functional form $A + B(ua)^2$, with confidence level 94\%.

It's interesting to compare to what would have been obtained if
we had used the same data but instead plotted the uncorrected
\begin {equation}
   {\Delta\langle\phiz^2\rangle_\c \over u_0}
   = {1\over u_0} \left[ \langle\phiz^2\rangle
                            - {\Sigma\over 2\pi a} \right]
\end {equation}
vs. $u_0 a$,
where $\phiz$ and $u_0$ are the bare lattice fields and coupling
of (\ref{eq:S0}).  This uncorrected data is represented by the
diamonds in Fig.\ \ref{fig:Lu144}.  One clearly sees the $O(a)$ corrections.
We should make clear that this is still $Lu=144$ data and is not
$Lu_0=144$ data, which would have required additional simulations.

\begin {figure}
\vbox{
   \begin {center}
      \epsfig{file=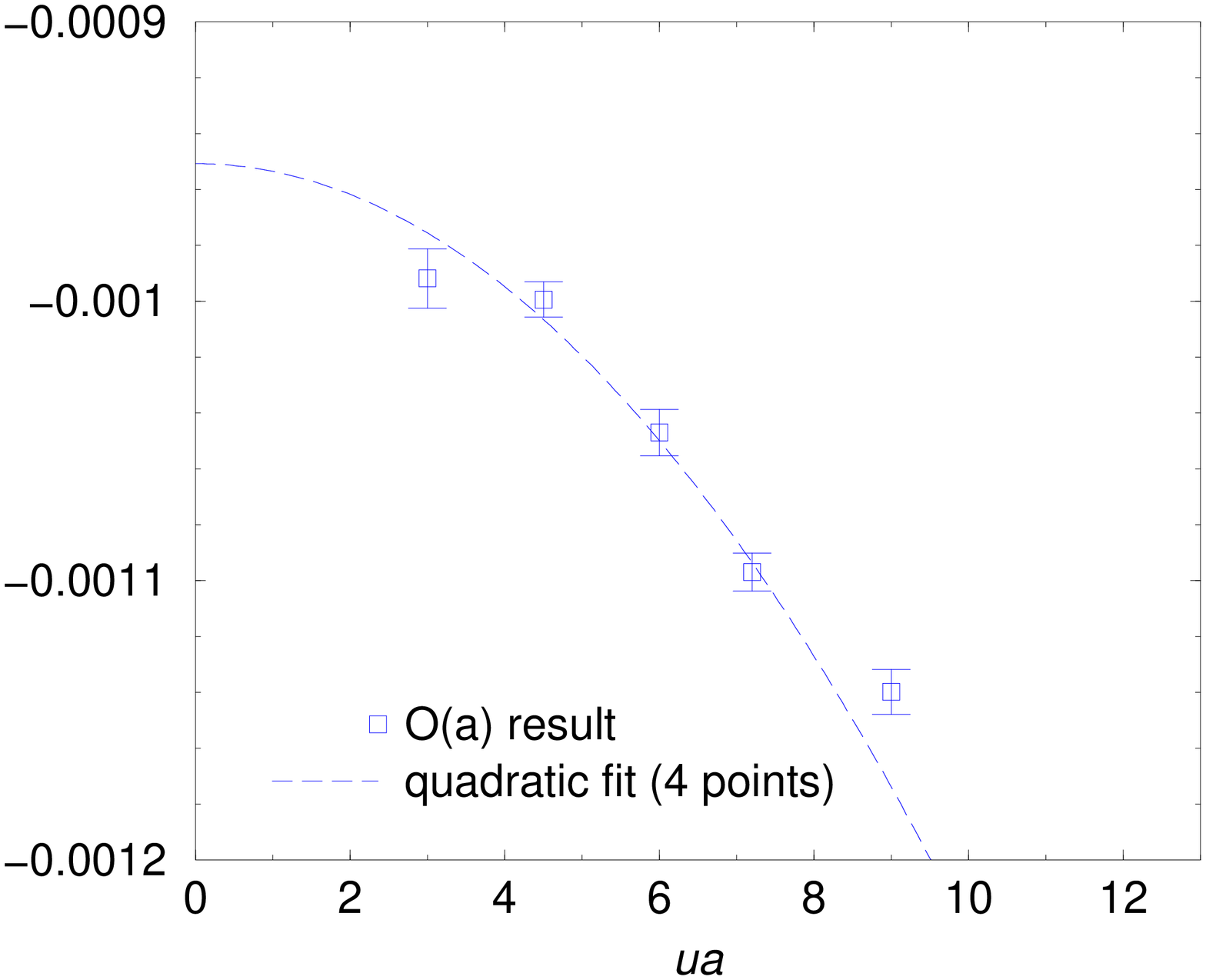,scale=.40,angle=-90}
   \end {center}
   \caption{
       Results for $\Delta\phiphic/u$ vs.\ $ua$ at $Lu=576$.
       The line is a fit of
       $A + B(ua)^2$ to all but the rightmost data point.
       \label{fig:Lu576}
   }
}
\end {figure}

\begin {figure}
\vbox{
   \begin {center}
      \epsfig{file=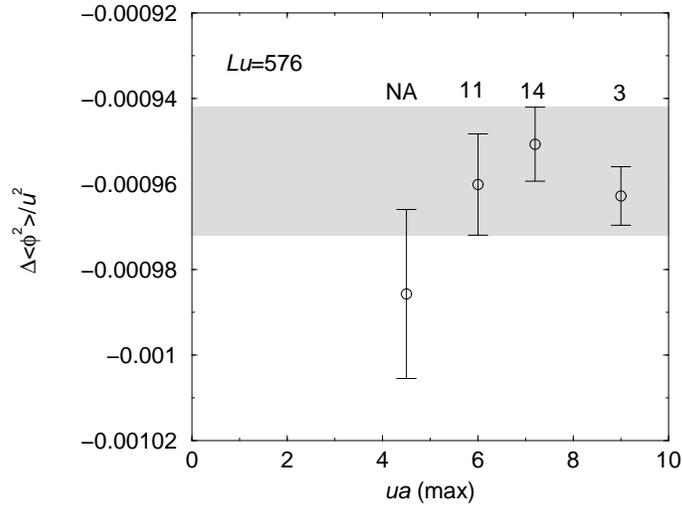,scale=.40,angle=-90}
   \end {center}
   \caption{
       Extrapolations of the data of Fig.\ \ref{fig:Lu576} to $ua=0$ fitting
       to the form $A + B(ua)^2$.
       The horizontal axis shows the maximum value of $ua$ used for each
       fit.
       Confidence levels are written as described for Fig.\ \ref{fig:ua6fits1}.
       \label{fig:Lu576fit}
   }
}
\end {figure}

Fig.\ \ref{fig:Lu576} shows the $ua$ dependence at a reasonably large
physical system size of $Lu=576$.
We show extrapolations of
the $ua \to 0$ limit in fig.\ \ref{fig:Lu576fit}.
Here, the 10\% confidence levels of fits to $A + B (ua)^2$ are less
spectacular than the $Lu=144$ data, though not unreasonable.
Because of the lower confidence levels, we have been a little more
conservative in our error estimate.  We take as our result for
the $ua \to 0$ limit the shaded region of fig.\ \ref{fig:Lu576fit},
which has been chosen to cover both the 10--15\% confidence level fits:
\begin {equation}
   \left[\Delta\phiphic\over u^2\right]_{Lu=576} = -0.000957(15) .
\label{eq:uax576}
\end {equation}
Combining this with the finite volume correction (\ref{eq:Lcorrect576}),
and adding errors in quadrature,
we arrive at our final value (\ref{eq:finalphisqr})
for the infinite-volume continuum limit.
Note that the dominant error in this estimate comes from
the $ua\to0$ extrapolation.

\subsection {Extrapolating $r_\c$ to $ua \to 0$}

\begin {figure}
\vbox{
   \begin {center}
      \epsfig{file=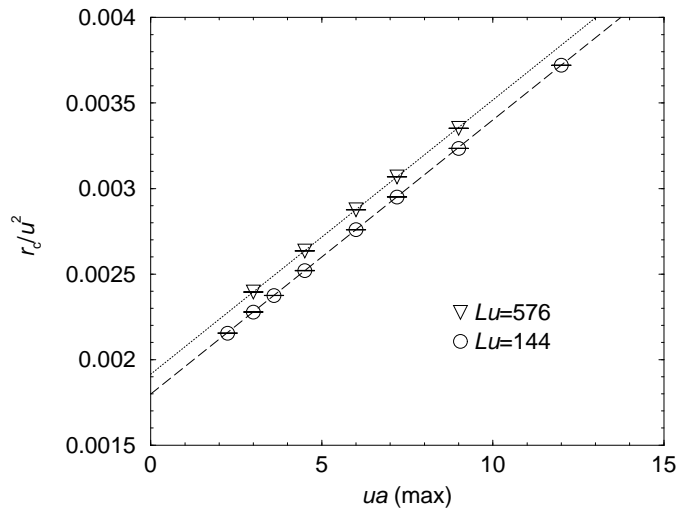,scale=.40,angle=-90}
   \end {center}
   \caption{
       Results for $r_\c/u^2$ vs.\ $ua$ at $Lu=144$
       and $Lu=576$
       (defined in \MSbar\ renormalization at renormalization
       scale $\bar\mu=u/3$).
       The lines are linear fits.
       \label{fig:musqr}
   }
}
\end {figure}

\begin {figure}
\vbox{
   \begin {center}
      \epsfig{file=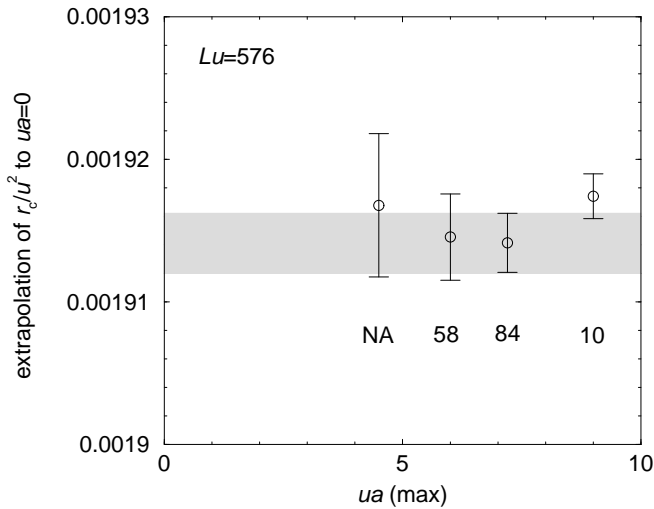,scale=.40,angle=-90}
   \end {center}
   \caption{
       Linear extrapolations of the $Lu=576$ data of Fig.\ \ref{fig:musqr}
       to $ua=0$.
       The horizontal axis shows the maximum value of $ua$ used for each
       fit.
       Confidence levels are written as described for Fig.\ \ref{fig:ua6fits1}.
       \label{fig:Lu576musqrfit}
   }
}
\end {figure}

The circles in Fig.\ \ref{fig:musqr} show the dependence of $r_\c$ on $ua$ at
the medium system size of $Lu=144$, where we can simulate down to
relatively small values of $ua$.  The data fits well to a linear
dependence on $ua$, with the fit shown to the first six points
in the figure having a 38\% confidence level.
The triangles in the same figure show similar dependence
for the reasonably large system size of $Lu=576$.
The corresponding extrapolations
to $ua=0$ based on linear fits are shown in fig.\ \ref{fig:Lu576musqrfit}.
We take the $ua=0$ extrapolation to be
\begin {equation}
   \left[ r_\c(u/3)\over u^2 \right]_{Lu=576}
     = 0.0019141(21) ,
\label {eq:uax576rc}
\end {equation}
as shown by the shaded region of the figure.
Combining with our estimate (\ref{eq:Lcorrect576r}) of the finite size
error at $Lu=576$ gives our final result (\ref{eq:finalrc})
for the infinite-volume
continuum limit.
Again, the dominant error comes from the $ua \to 0$ extrapolation.

\subsection {Interdependence of \boldmath$Lu\to\infty$ and \boldmath$ua\to0$
             extrapolations}
\label {sec:interdependence}

In this paper, we have treated the extrapolation of our
finite volume correction, taken from $ua{=}6$ data, as independent from
our continuum extrapolation, taken from $Lu{=}576$ data.
Consider the case of $\Delta\phiphic/u$.
In principle, the two extrapolations are not completely independent
because the coefficient $B$ in the form $A + B L^{-d/2}$
used for our large $L$ extrapolation is not universal.  Its value
will have some dependence on $ua$.  For our procedure, we would like
to know the value of $B$ at $ua=0$, to obtain
the large $L$ correction (\ref{eq:Lcorrect576}) at $ua=0$.
Instead, we have the large $L$ correction at $ua=6$, which will be
slightly different.
In numerical terms, we will have underestimated our systematic error if
$B[ua{=}6]$ differs from $B[ua{=}0]$ by $\sim 7$\%, which is the
final error on $\Delta\phiphic/u$ quoted in (\ref{eq:finalphisqr})
relative to the size of the finite volume correction
(\ref{eq:Lcorrect576}).

How does the coefficient $B$
depend on $ua$ as $ua \to 0$?  Because of our improvements to the
action and to the operator $\phi^2$, the answer is that small $ua$
corrections to $B$ vanish as $(ua)^3$.
To see this, note that the only quantities we have {\em not} matched
through $O((ua)^2)$ are $r$ and $\delta \phi^2$.  However, neither of
these contribute to $B$.  $\delta \phi^2$ is a constant and therefore
$Lu$ independent, and our procedure tunes to the critical value of (the
continuum parameter) $r$ whether or not the lattice parameter is
correctly matched to the continuum one.  The residual difference between
lattice and continuum $r$ does appear in Eq.~(\ref{eq:dphimaster}), but,
because $r-r_c$ vanishes as $L^{-\omega{-}d/2}$, this does not contribute to
$B$ either.
We should not expect such a high order correction to be of any
consequence.  

There is an alternate way of fitting our data which makes it easier to
check this.  We can fit all data at large $Lu$ and relatively small $ua$
to a form which allows both spacing and volume dependence,
\begin{equation}
{\Delta\phiphic\over u} = A + B (Lu)^{-y_t} + C (ua)^{2} \, .
\end{equation}
Based on our fits at fixed $Lu$ and fixed $ua$, we expect this form is
sufficient if we use only data
with $Lu \geq 384$ and $ua \leq 8$.  Fitting all such data gives
$\Delta \phiphic/u = -0.001194(8)$ with
$\chi^2 / \mbox{d.o.f.} = 15.5/8$ (5\%
confidence level).  The result is consistent with our quoted result, and
has smaller errors.  The quality of the fit is not very good, which is why
we quote the larger error bars of our other analysis.  However,
adding a term which allows $B$ to vary with $ua$ (that is, a 
$ua$ and $Lu$ dependent term) does not improve the
quality of the fit.  Adding a
new term $D (Lu)^{-y_t} (ua)^{m}$ and doing a 4 parameter fit changes
$\chi^2$ by less than 1 for $m=2$ or $m=3$; so the data show no evidence
that such a term is large enough to be important.

We do not understand why the fit to all data has such a large $\chi^2$
($\chi^2/\mbox{d.o.f} = 15.3/8$).
The three lowest $ua$ data points are the main
outliers; the two $ua=3$ points are each at 2$\sigma$.  However, the
$(Lu,ua)=(384,3)$ point is off in the opposite direction as the
$(Lu,ua)=(576,3)$ and $(Lu,ua)=(384,4)$ points, so there is no systematic
trend in the residuals and we have no reason to believe that some
additional lattice spacing or volume dependent term is missing in our
fitting procedure.

Finally, there is a minor, independent issue in our main analysis.  
The statistical part of our errors in our original infinite volume and
continuum extrapolations are correlated, since both of these
extrapolations included the $(Lu,ua)=(576,6)$
data point.  However, since the error from the continuum extrapolation
dominated that from the infinite volume extrapolation, any
statistical cross-correlation between those errors will not have
a significant effect.


\section* {ACKNOWLEDGMENTS}

We are indebted to John Cardy for outlining to us the argument
of appendix \ref{app:log}.
We also thank Alan Sokal for useful discussions.
This work was supported by the U.S. Department
of Energy under Grant Nos.\ DE-FG03-96ER40956
and DE-FG02-97ER41027.

\clearpage
\appendix


\section {Tabulated data}
\label {app:table}

\begin {center}
\setlength{\tabcolsep}{8pt}
\begin {tabular}{rlrllcr}                          \hline
$Lu$ & $ua$ & $L/a$ & \multicolumn{1}{c}{$\Delta\phiphic/u$}
  & \multicolumn{1}{c}{$r_\c/u^2$}
  & $\tau_{\rm decorr}$
  & $N_{\rm sweeps}/2\tau$
  \\  \hline
  8 & ~1 & 8 & ~0.2325(58) &-0.02992(98)              &  0.6 & 16868       \\  
  24 & ~1 & 24 & ~0.0400(28) &-0.00207(37)            &  0.5 & 1070        \\  
  24 & ~3 & 8 & ~0.0420(21) &-0.00249(26)             &  0.5 & 11470       \\  
  36 & ~3 & 12 & ~0.02164(96) &~0.00027(13)           &  0.5 & 38070       \\  
  48 & ~3 & 16 & ~0.012322(95) &~0.001233(12)         &  0.6 & 231566      \\  
  72 & ~3 & 24 & ~0.00575(15) &~0.001906(25)          &  0.7 & 9993        \\  
  96 & ~3 & 32 & ~0.003212(65) &~0.0021036(79)        &  0.8 & 26406       \\  
  96 & ~6 & 16 & ~0.0030836(86) &~0.0025981(12)       &  0.8 & 1235650     \\  
  144 & ~2.25 & 64 & ~0.001039(18) &~0.0021542(24)    &  1.3 & 205420      \\  
  144 & ~3 & 48 & ~0.0010254(98) &~0.0022783(16)      &  1.4 & 158458      \\  
  144 & ~3.6 & 40 & ~0.0010253(83) &~0.00237477(99)   &  1.1 & 373163      \\  
  144 & ~4.5 & 32 & ~0.0010267(97) &~0.0025199(13)    &  1.2 & 370142      \\  
  144 & ~6 & 24 & ~0.0010083(44) &~0.00275922(62)     &  1.1 & 940826      \\  
  144 & ~7.2 & 20 & ~0.0009940(40) &~0.00295088(70)   &  1.0 & 863928      \\  
  144 & ~9 & 16 & ~0.0009222(43) &~0.00323454(58)     &  1.1 & 1213680     \\  
  144 &  12 & 12 & ~0.0009256(43) &~0.00372103(51)    &  0.7 & 1844310     \\  
  192 & ~6 & 32 & ~0.0001118(88) &~0.0028151(14)      &  1.7 & 192227      \\  
  288 & ~3 & 96 & -0.0004299(90) &~0.0023640(13)      &  5.7 & 21604       \\  
  288 & ~6 & 48 & -0.0005625(76) &~0.0028531(10)      &  3.4 & 50516       \\  
  384 & ~3 & 128 & -0.0007484(57) &~0.00238397(94)    &  3.7/5.7 & 69038   \\  
  384 & ~4 & 96 & -0.000799(11) &~0.0025498(15)       &  7.4 & 12316       \\  
  384 & ~4.8 & 80 & -0.0008055(80) &~0.0026760(14)    &  6.3 & 18797       \\  
  384 & ~6 & 64 & -0.0008455(71) &~0.0028682(10)      &  5.7 & 14591       \\  
  384 & ~8 & 48 & -0.0009131(78) &~0.0031875(12)      &  3.7 & 20958       \\  
  384 &  12 & 32 & -0.01217(18) &~0.0038358(19)       &  3.6 & 6021        \\  
  576 & ~3 & 192 & -0.000992(11) &~0.0023957(16)      &  17.5/8.5 & 4458   \\  
  576 & ~4.5 & 128 & -0.0009993(64) &~0.00263522(80)  & 12.5 & 12212       \\  
  576 & ~6 & 96 & -0.0010470(83) &~0.0028761(11)      &  7.4/7.9 & 7196    \\  
  576 & ~7.2 & 80 & -0.0010968(68) &~0.00306843(97)   & 10.8/6.2 & 13493   \\  
  576 & ~9 & 64 & -0.0011398(80) &~0.0033525(11)      & 10.5 & 4007        \\  
  768 & ~6 & 128 & -0.001142(11) &~0.0028802(13)      & 14.3/13.1 & 3074   \\  
  1152 & ~6 & 192 & -0.0011999(86) &~0.00288154(80)   &  7.4/18.5/10.1&6342\\

\hline
\end {tabular}
\end {center}

\bigskip

To give a rough idea of the size of our data sets, we have listed above a
nominal decorrelation time $\tau$ for each simulation, along with the amount of
data we have in units of $2\tau$.
In our convention, $\tau=0.5$ represents completely uncorrelated data.
Our decorrelation times are in units of sweeps, where one ``sweep'' consists
of both a heatbath sweep and a multi-grid update.
More than one time is shown in cases where simulations were made at
different values of $r$ (before reweighting).

Our nominal decorrelation time is the largest integrated decorrelation time
of the various expectations required in the computation of the Binder cumulant
and of $\langle\phi^2\rangle$.  These include the integrated decorrelation
times associated with measurements of $P$, $P \phi^2$, $P \bar\phi^2$,
and $P \bar\phi^4$ (and also times associated with cross-correlations
between these), where
\begin {equation}
   P \equiv \exp\left(-\half V\,\delta r\,\overline{\phi^2}\right) 
\end {equation}
is the canonical reweighting factor, as in (\ref{eq:reweight}).
We found in all cases that the longest decorrelation time
was that associated with $\bar\phi^2$.
The integrated decorrelation time for a single operator is given by
\begin {equation}
   \tau \equiv {1\over2} + \sum_{n=1}^\infty {C(n)\over C(0)} ,
\label {eq:tau}
\end {equation}
where
\begin {equation}
   C(n) =
     {1\over(N-n)} \sum\limits_{i=1}^{N-n} O_i O_{i+n}
     - \left({1\over N}\sum\limits_{i=1}^N O_i\right)^2 ,
\end {equation}
is the auto-correlation function associated with the desired operator $O$.
In practice, the sum in (\ref{eq:tau}) must be cut off because of
degrading statistics, and we cut it off when $C(n)/C(0)$ first drops below
$0.05$.


\section {Matching continuum and lattice theories}
\label {app:match}

\subsection {General discussion}

In this appendix we discuss the $O(a)$ and $O(a^2)$ improvement of
three-dimensional scalar field theory on the lattice.
For the sake of generality, and because it is not any harder, we
will discuss O($N$) scalar field theory of $N$ real fields
$\phi = (\phi_1,\phi_2, \cdots \phi_N)$.
The case of interest to the present work is $N=2$.
As in (\ref{eq:Slat}), the lattice Lagrangian is defined to be
\begin {equation}
   {\cal L} = a^3 \sum_\x \left[
      {Z_\phi\over2} \sum_{i=1}^N
           (- \phi_i \grad_\lat^2 \phi_i )
      + {Z_r (r+\delta r)\over2} \phi^2
      + {u+\delta u\over4!} (\phi^2)^2 
   \right] ,
\end {equation}
where
\begin {equation}
   \phi^2 \equiv \sum_{i=1}^N \phi_i^2 ,
\end {equation}
and $\phi$, $r$, and $u$ are just the (UV renormalized) continuum fields
and parameters
in lattice units.  So $\phi_\lat = a^{1/2} \phi_\cont$, $u_\lat = a u_\cont$,
and $r_\lat = a^2 r_\cont$.  Of these,
$r_\cont$ is the only continuum parameter that
requires UV renormalization and should be understood as renormalized with
dimensional regularization and the \MSbar\ scheme.
In this appendix, we will calculate, to a given order in lattice spacing,
the necessary counter-terms $Z_\phi$, $Z_r$,
$\delta r$, and $\delta u$ required to implement this correspondence
between continuum and lattice variables.
To match the lattice and continuum theories to high orders in $a$, we
would need to include other operators in our lattice theory, such as
$\phi^6$, $\phi^2 |\grad \phi|^2$, and so forth.  However, these will
turn out to be unnecessary at the order to which we will work.

A given local lattice action will never perfectly reproduce the continuum
action.  For small $ua$, the error in how a given lattice action treats
physics at the distance scale $a$ can be computed and compensated for by
perturbative calculations.  The discrepancy in how a given lattice action,
with given parameters, treats the non-perturbative physics at the
distance scale $1/u$ cannot.  It is therefore important that the
lattice action be close enough to the continuum action that errors at
the scale $1/u$ are higher order in $a$ than whatever is desired.
In order to improve our simulations and measurement of
$\Delta\phiphic$ to $O(a)$ accuracy, it is necessary to
(a) match the lattice action and parameters
so that, at the scale $1/u$, it reproduces
the physics of the continuum action up to and including $O(a)$;
and (b) match the lattice and continuum definitions
of operator $\phi^2$ to the same order, so that the measurements of
$\Delta\phiphic$ will match up.
Actually, the first requirement is slightly overstated.  To measure
$\Delta\phiphic$ to $O(a)$, it is not necessary to match
the lattice and continuum parameters $r$ to that order.  That's because,
in the simulations, we will simply vary the coefficient of
$\phi^2$ in the action until we find the transition -- we don't need
to know its relation to $r_\cont$ to do this.

In this appendix, we will match $r_\cont$ to the lattice just to $O(a^0)$.
This will make possible a determination of the continuum value of the
critical value $r_\c$, but $O(a)$ errors will remain and must be removed
by extrapolation to the continuum limit.
We will match the continuum and lattice definitions of the operator $\phi^2$
through $O(a)$, so that we can make an $O(a)$ improved calculation of
$\phiphic$.  We will also match
the lattice action through $O(a^2)$ [except for the matching of $r$
just discussed], rather than simply $O(a)$, because it is not that
difficult [compared to the $O(a)$ matching of $\phi^2$] and should
improve the $ua$ dependence of infinite-volume extrapolations of
our data, as discussed in section \ref{sec:interdependence}.

\begin{figure}
  \centerline{\epsfxsize=10.5cm\epsfbox{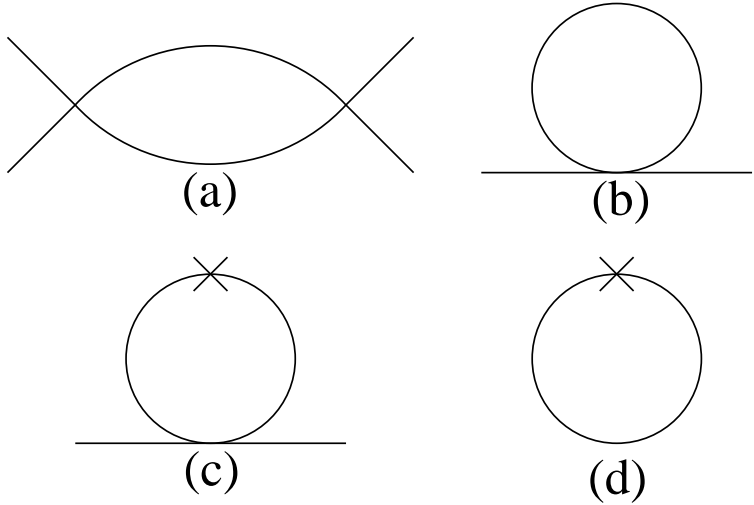}}
  \caption{
     \label{one_loop_graphs}
     The one loop graphs needed for the
     renormalization.  A cross represents a $\phi^2$ insertion, which in
     turn could represent either (i) a perturbative insertion of the
     $r\phi^2$ term
     of the action or (ii) the $\phi^2$ operator associated with calculating
     the expectation $\langle\phi^2\rangle$.
  }
\end{figure}

At tree level, the infrared behavior of the lattice and continuum
theories are the same up to corrections suppressed by at least $a^2$;
the power of $a$ can be higher if we choose the lattice $\nabla^2$
appropriately.  However there are ``radiative'' corrections induced in
the IR physics by the nonlinearity of the theory, together with the UV
difference between the lattice and continuum theories.  The diagrams of
Fig. \ref{one_loop_graphs}, for instance, differ on the lattice from
their continuum values because the dispersion relations of the scalar
fields differ in the UV, and the loop momentum integration integrates
over this region.  These effects are suppressed by powers of the couplings;
the radiative correction to the scalar 4 point function, for instance,
clearly depends on $u^2$.  On dimensional grounds, the difference
between lattice and continuum values of this diagram 
must go as $u^2 a$, so these
diagrams lead to $O(a)$ differences between the lattice and continuum
theories.  The difference between lattice and
continuum values of these diagrams can be removed by a renormalization
of the parameters of the lattice theory.

\begin {figure}
\vbox{
   \begin {center}
      \epsfig{file=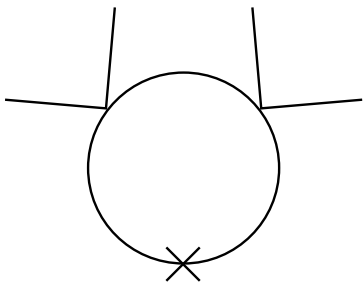,scale=.50}
   \end {center}
   \caption{
       A diagram contributing to the mixing of the $\phi^2$ and $\phi^4$
       operators.
       \label{fig:mix}
   }
}
\end {figure}

As discussed in the main text, we will be interested not only in how to
renormalize the parameters of the action but also in how to
translate expectation values of the operator $\phi^2$
between the lattice and continuum.
It will be convenient to talk directly about the operator $\phi^2$, which
is associated with UV divergences in the continuum, rather than
$\Delta\langle\phi^2\rangle$, which is not.
We will define $\phi^2$ in the continuum also using \MSbar\ renormalization.
In general, operators with the same symmetry can mix under renormalization,
and the lattice operator $\phi^2$ will correspond to some superposition
of the unit continuum operator, the renormalized $\phi^2$ continuum operator,
and higher-dimensional renormalized continuum operators such as $\phi^4$:
\begin {equation}
   a^{-1} (\phi^2)_\lat
         = c_0 + c_2 (\phi^2)_\cont + c_4 (\phi^4)_\cont + \cdots .
\label {eq:oprmix}
\end {equation}
However, our particular interest is in expectation values at the transition.
For this application, the effects of higher-and-higher dimensional
operators on the right-hand side of (\ref{eq:oprmix})
become suppressed by more and more powers of $ua$.
For example, $c_2$ is O(1), and the expectation value
of the renormalized continuum operator $\phi^2$
is, by dimensional analysis, $O(u)$ at the transition.
So the $c_2 (\phi^2)_\cont$ term contributes
$O(u)$ to the expectation.  In contrast, the lowest-order diagram that
contributes to mixing between $\phi^2$ and $\phi^4$ is shown in
Fig.\ \ref{fig:mix}
and gives $c_4 = O(u^2 a^3)$, where the $u^2$ counts the
two vertices in Fig.\ \ref{fig:mix} and the $a^3$ then follows by dimensional
analysis.  But, also by dimensional analysis, the renormalized continuum
expectation $\langle\phi^4\rangle$ is $O(u^2)$.  So
the $c_4 (\phi^4)_\cont$ term contributes $O(u^4 a^3)$ to the
expectation, down by three powers of $ua$ from $c_2 (\phi^2)_\cont$.
Our goal will be to compute the $O(a)$ corrections to
$\Delta\phiphic$, for which it is therefore adequate to compute
just $c_0$ and $c_2$ above.

The diagrams which contribute to
$c_2$ are the same as those which contribute to the multiplicative
renormalization of the coefficient $r$ of $\phi^2$ in the action,
and $1/c_2$ is the same as the $Z_r$ introduced previously.
Rearranging the terms in (\ref{eq:oprmix}), we will write
\begin {equation}
   a (\phi^2)_\cont = Z_r (\phi^2)_\lat - \delta\phi^2
           + O\blparen(ua)^4\brparen ,
\end {equation}
where $\delta\phi^2$ represents a $c$-number ($ac_0/c_2)$
parameterizing mixing with the unit operator, which we shall calculate.

We will see later that, for the purpose of calculating
the $O(a)$ corrections to $\Delta\phiphic$,
all that is strictly required is one-loop results for
$Z_\phi$, $Z_r$, $\delta u$, and $\delta r$, and three-loop results for
$\delta\phi^2$.
To determine $r_c$ [just at $O(a^0)$]
requires tree results for $Z_\phi$, $\delta u$, and
$Z_r$, and two-loop results for $\delta r$.
If one made a three-loop computation of $\delta r$ (which we have not),
one could obtain the $O(a)$ corrections for $r_c$ as well.
To match the action through $O(a^2)$, except for $r$, one wants a
two-loop determination of $Z_\phi$ and $\delta u$.
As discussed in section \ref{sec:interdependence}, it will also be
useful to have a two-loop result for $Z_r$.

Now we turn to the calculations.
For the remainder of this appendix, we will work exclusively in lattice
units ($a=1)$.


\subsection {One-loop results}

A one loop renormalization calculation will determine the $O(a)$
contributions to the quantities $Z_{\phi}$, $\delta u$, and $Z_r$,
and will find the $O(1/a)$ contributions to $\delta m^2$
and $\delta \langle \phi^2 \rangle$.  The details of the power counting
used here can be found in \cite{Oapaper}.
In the small lattice spacing limit, the momentum scale $r^{1/2}$ associated
with the parameter $r$ is small compared to the scale $1/a$ where the
lattice and continuum theories differ.  So, for the specific
purpose of a calculation to match the lattice and continuum theories,
the $r\phi^2$ term in the action (as well as the $u \phi^4$ term)
may be treated perturbatively.

The required graphs are shown in
Figure \ref{one_loop_graphs}.  Evaluating the graphs requires choosing a
lattice Laplacian, and for completeness 
we will consider both the unimproved (\ref{eq:lapU})
and improved (\ref{eq:lapI}) choices described in the main text.
The evaluation of the one loop graphs requires two integrals:
\begin {eqnarray}
\label{sigma_is}
  \frac{\Sigma}{4 \pi} & \equiv & \int_{\BZ} \frac{d^3 k}{(2\pi)^3} \, 
        \frac{1}{\tilde{k}^2} \, ,
\\
\label{xi_is}
  \frac{\xi}{4 \pi} & \equiv & \int_{\BZ} \frac{d^3 k}{(2\pi)^3} \,
        \frac{1}{(\tilde{k}^2)^2} - \int_{\Re^3} \frac{d^3 k}{(2\pi)^3} 
        \, \frac{1}{k^4} \, .
\end {eqnarray}
Here we use the shorthand $\BZ$ to mean $k$ lies within the Brillouin
zone, meaning each $k_i \in [-\pi,\pi]$.  The notation $\tilde{k}^2$ is
introduced in Eq.~(\ref{eq:ktw_def}).
The integrals which determine
$\xi$ are each IR singular and some regulation is implied, for instance
adding $m^2$ to both $k^2$ and $\tilde{k}^2$ and taking the limit as
$m^2 \rightarrow 0$.  The numerical values of the integrals, accurate to 
$\pm 1$ in the last digit, are
\footnote{
   An analytic result \cite{watson} for $\Sigma_\U$ is
   ${8\over\pi}(18+12\sqrt{2}-10\sqrt{3}-7\sqrt{6})
   \bigl[{\bf K}\blparen(2-\sqrt{3})^2
           (\sqrt{3}-\sqrt{2})^2\brparen\bigr]^2$,
   where ${\bf K}$ is the complete elliptic integral
   of the first kind.
}
\begin {eqnarray}
  \Sigma_{\rm U} & = & 3.17591153562522 \, , \qquad \hspace{0.2em}
  \Sigma_{\rm I} =     2.75238391130752 \, , \nonumber \\
  \xi_{\rm U} & = & 0.152859324966101 \, , \qquad
  \xi_{\rm I} =    -0.083647053040968 \, .
\end {eqnarray}
Note that the sign of $\xi$ depends on whether we use an improved
lattice Laplacian.  This is possible because $\xi$ represents the
difference of a graph between lattice and continuum theories.  The
lattice contribution is larger inside the Brillouin zone, but the
continuum integral receives contributions from outside the zone as well;
the sign depends on which effect is larger.

At one loop the renormalizations are (in lattice units)
\begin {eqnarray}
  \delta u_{1l} &=& {(N+8)\over 6} \, {\xi\over4\pi}  \, u^2 , \\
  Z_{\phi,1l}-1 &=& 0 , \\
  Z_{r,1l} - 1 &=& {(N+2)\over 6} \, {\xi\over4\pi}  \, u , \\
  \delta r_{1l} &=& - {(N+2)\over6} \, {\Sigma\over4\pi} u \,, \\
  \delta \phi^2_{1l} &=& N \, {\Sigma\over4 \pi} \, .
\label{dphi_1l}
\end {eqnarray}


\subsection{Two-loop results}

\begin{figure}[t]
  \centerline{\epsfxsize=10.5cm\epsfbox{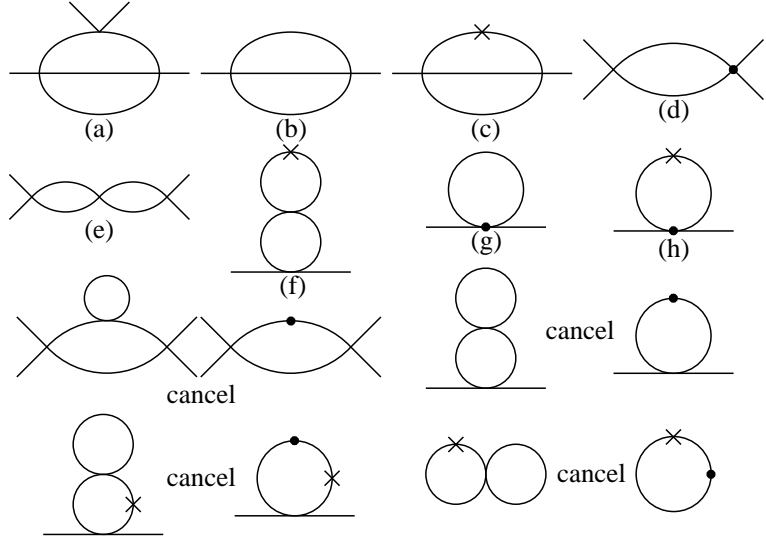}}
  \vspace{0.1in}
  \caption{
    \label{two_loop_graphs}
    All required two loop graphs and one
    loop graphs with one loop counterterm insertions, shown as heavy dots on
    lines (mass counterterms) or at vertices (coupling counterterms).   The
    last eight graphs cancel in pairs.  Diagrams (a), (d), and (e) are not
    separately IR convergent; diagram (d) must be distributed between the
    other two to produce IR convergent integrals.
  }
\end{figure}

For the renormalizations $Z_\phi$ and $Z_r$, it makes no sense to carry
the renormalization to two loops unless we use the improved lattice
Laplacian, because already at tree level the unimproved Laplacian
gives $O(a^2)$ level errors in the propagator.
The two loop results require several more
graphs as well as the inclusion in one-loop graphs of one-loop mass and
coupling counterterms.  (See Figure \ref{two_loop_graphs}.)
Four more
integrals are needed, which we will evaluate in a moment.
The complete 2-loop result for general $N$ is
\begin {eqnarray}
  \delta u_{2l} - \delta u_{1l} &=& 
    \left[
       {(N^2+6N+20)\over 36}\left(\xi\over4\pi\right)^2
       - {(5N+22)\over9}\, {C_1\over(4\pi)^2}
    \right] u^3 ,
\\
  Z_{\phi,2l} - 1 &=&
    {(N+2)\over 18} \, {C_2\over (4\pi)^2} \, u^2 ,
\\
  Z_{r,2l} - Z_{r,1l} &=&
    \left[\left(N+2\over6\right)^2\left(\xi\over4\pi\right)^2
           - {(N+2)\over6} \, {C_1\over(4\pi)^2} \right] u^2 ,
\\
  \delta r_{2l} - \delta r_{1l} &=&
    {(N+2)\over18 (4\pi)^2} \left[
        C_3 + \ln\left(6\over\bar\mu\right) - 3 \Sigma \xi
    \right] u^2 ,
\\
  \delta\phi^2_{2l} - \delta\phi^2_{1l} &=&
     {N(N+2)\over 6} \, {\Sigma \xi \over (4\pi)^2} \, u .
\end {eqnarray}
The three numerical constants $C_1$, $C_2$, and $C_3$ are
given for the improved Laplacian in (\ref{eq:consts}).
Only $C_3$ can be usefully defined for
the unimproved case,%
\footnote{
   In ref.\ \cite{FKRS}, this constant is called
   $\zeta$.
}
where it is $C_{3,{\rm U}} = .08848010$.

Now we detail the calculation of the constants $C_1$ through $C_3$.
We will use the following shorthands:  $\int_{k,\BZ}$ means
$\int d^3 k/(2\pi)^3$, with range the Brillouin zone $[-\pi,\pi]^3$; whereas
$\int_{k,\Re^3}$ is the same but integrated over all 3-space.  Further,
we define the following integrals, which will come up repeatedly:
\begin{eqnarray}
\IL(p) & = & \int_{q,\BZ} \frac{1}{\tilde{q}^2 
        ( \negsp \widetilde{ \: \; p+q \: \;} \negsp )^2} \, , \\
\IC(p) & = & \int_{q,\Re^3} \frac{1}{q^2(p+q)^2} 
        = \frac{1}{8|p|} \, .
\end{eqnarray}

We begin with the two-loop vertex correction, graph (a) of Figure
\ref{two_loop_graphs}.  The required integral, including the
appropriate amount of the one loop counterterm graph (d), is
\begin {equation}
\frac{C_1}{(4\pi)^2} = 
        \int_{k,\BZ} \frac{1}{(\tilde{k}^2)^2} \left\{\IL(k) 
        - \frac{\xi}{4 \pi} \right\} 
        - \int_{k,\Re^3} \frac{1}{k^4} \IC(k) \, .
\end {equation}
We re-arrange the original integral into three parts,
\begin {equation}
\int_{k,\BZ} \frac{1}{(\tilde{k}^2)^2} \left\{ \IL(k)
         - \frac{1}{8k} -
        \frac{\xi}{4 \pi} \right\} + \int_{k,\BZ} \frac{1}{8k} \left( 
        \frac{1}{(\tilde{k}^2)^2} - \frac{1}{k^4} \right) 
        - \int_{k,\Re^3-\BZ} \frac{1}{8 k^5} \, .
\label {eq:C1aa}
\end {equation}
All three of the above integrals are convergent, provided we use the improved
Laplacian.
The first integral is IR well behaved because the two counterterms
cancel $\IL(k)$ up to a $k^2$ correction, which in the small $k$
limit is $\IL(k)-(1/8k)-(\xi/4\pi) \rightarrow
.0125438 \,  k^2 / 4\pi$.

To get accurate numerical answers, we
perform all 3-D integrals by
quadratures.  Dealing with the double poles which appear in
$\IL$ is touchy, and requires adaptive mesh refinement techniques.
We improve
the precision of each final result by repeating the full
integration at several spacings and extrapolating (Richardson
extrapolation).  
The first integral in (\ref{eq:C1aa}) gives $.0360003/16\pi^2$,
and the second gives $.054568958/16 \pi^2$.
The last integral, over $\Re^3-\BZ$, can
be re-arranged into
\begin {equation}
-\frac{3}{16 \pi^5} \int_0^1 dx \int_0^1 dy \frac{1}{(1+x^2+y^2)^{5/2}}
         \, , 
\end {equation}
which can be performed very accurately and gives $-.035507296027
\cdots/16 \pi^2$.  These sum to give $C_1 = .0550612$.

Besides this graph, there is graph (e), which gives
\begin {equation}
\left( \int_{k,\BZ} \frac{1}{(\tilde{k}^2)^2} \right)^2 - 
        \left( \int_{k,\Re^3} \frac{1}{k^4} \right)^2 ,
\end {equation}
which is {\em not} IR convergent.  However, including -2 times the
counterterm diagram (d),
\begin {equation}
-2 \left( \int_{k,\BZ} \frac{1}{(\tilde{k}^2)^2} \right)
        \left(  \int_{k,\BZ} \frac{1}{(\tilde{k}^2)^2} - 
         \int_{k,\Re^3} \frac{1}{k^4} \right) ,
\end {equation}
gives $-(\xi/4\pi)^2$.  No new integrals are required.  It is a
nontrivial check on the calculation that the sum of the coefficients
arising from diagrams (a) and (e) precisely absorb diagram (d).

The next integral is the $O(p^2)$ contribution from the setting sun
diagram (b), 
\begin {eqnarray}
\frac{C_2}{(4 \pi)^2} = \lim_{p\rightarrow 0} \frac{1}{p^2}
        \Bigg\{ \int_{k,\BZ} \left( 
        \frac{1}{
        (\negsp \widetilde{ \; \: k{+}p \: \;} \negsp )^2
        } - \frac{1}{\tilde{k}^2} \right) 
        \IL(k)
        - \int_{k,\Re^3} \left( \frac{1}{(k{+}p)^2}-\frac{1}{k^2} \right)
        \IC(k) \Bigg\} .
\label {eq:C2aa}
\end {eqnarray}
The first trick is to note that 
\begin {equation}
\int_{k,\BZ} \left( \frac{1}{
        ( \negsp \widetilde{\; \: k{+}p \: \;} \negsp )^2
        } - \frac{1}{\tilde{k}^2} \right) = 0
\end {equation}
just by shifting the integration variable for the
first term.  So we may add an arbitrary constant to
$I_l(k)$ in (\ref{eq:C2aa}),
and we choose the constant $-\xi/4\pi$.  This will prevent IR divergences
in what follows.  We are now free to expand
$1/( \negsp  \widetilde{ \; \: k{+}p \: \;} \negsp )^2$
to second order in $p$.  After averaging over directions for $p$, we find 
\begin {equation}
\frac{1}{
        ( \negsp \widetilde{\: \; k{+}p \; \: } \negsp )^2
        } - \frac{1}{\tilde{k}^2} \to p^2 \left[
        \frac{\frac{1}{3} \sum_i \left( \frac{8 \sin k_i - 
        \sin 2 k_i}{3} \right)^2 }{(\tilde{k}^2)^3} - \frac{ \frac{1}{3} 
        \sum_i \left( \frac{4 \cos k_i - \cos 2k_i}{3} \right)}
        {(\tilde{k}^2)^2} \right] \equiv p^2 {\cal M}(k) .
\end {equation}
The equivalent expression in the continuum case is $p^2/3k^4$.
Re-arranging the terms a little, we can write
\begin {equation}
\frac{C_2}{16 \pi^2} = \frac{-1}{24} \int_{k,\Re^3-\BZ} \frac{1}{k^5}
        + \int_{k,\BZ} \frac{1}{8k} \left( {\cal M}(k) -
        \frac{1}{3k^4} \right)
        + \int_{k,\BZ} \! \! \! \! {\cal M}(k) 
        \left[\IL(k)-\frac{1}{8k}-\frac{\xi}{4 \pi} \right] .
\end {equation}
We have seen the first integral.  
The second gives $.0310757695/16 \pi^2$ and the
last gives $.0142016/16 \pi^2$; so $C_2 = .0334416$.

Next we must compute the $O(p^0)$ part of the setting sun diagram.
The continuum diagram is IR and UV log divergent, while the lattice
diagram is only IR log divergent.  It is convenient to IR regulate both
by introducing a mass on one line.  In this case the continuum integral
can be performed in $\overline{\rm MS}$, leaving a lattice integral
minus an analytically determined counterterm \cite{FKRS,LaineRajantie}.
Choosing to separate the renormalization dependence along with the same
finite constant as in the previous literature \cite{FKRS,LaineRajantie},
the constant $C_3$ is given by
\begin {equation}
\frac{C_3}{(4 \pi)^2} = \lim_{m \rightarrow 0} \left\{ \int_{k,\BZ} 
        \frac{1}{\tilde{k}^2 + m^2} \IL(k)
        - \frac{1}{16 \pi^2} \left[ \frac{1}{2} 
        + \ln \frac{6}{m} \right] \right\} .
\end {equation}
The problem with this expression is the logarithm.  To remove it,
we add and subtract $\IC(k)=1/8k$ to $\IL(k)$.  The integral
\begin {equation}
\int_{k,\BZ} \frac{1}{\tilde{k}^2 + m^2} \left[ 
        \IL(k) - \frac{1}{8k} \right]
\end {equation}
is IR convergent, and the $m \rightarrow 0$ limit may be taken
immediately.  It evaluates to $-.06858432/16 \pi^2$, unless we use the
unimproved lattice Laplacian, in which case it is $.60953343 / 16
\pi^2$.  We re-arrange the remaining terms to be
\begin {equation}
\int_{k,\BZ} \left( \frac{1}{\tilde{k}^2 + m^2} - \frac{1}{k^2 + m^2} 
        \right) \frac{1}{8k} + \int_{k,\BZ} \frac{1}{8k(k^2+m^2)} - 
        \frac{1}{16 \pi^2} \left[ \frac{1}{2} + \ln \frac{6}{m} \right] 
        .
\end {equation}
Again, for the first integral the $m \rightarrow 0$ limit may be taken
immediately, and the numerical value is $.161799607 / 16 \pi^2$, or
$.43364112015 / 16 \pi^2$ if we use the unimproved Laplacian.  For the
last integral, we cut the integration region into the ball of radius
$\pi$ and the region within the Brillouin zone but outside the ball:
\begin {equation}
\int_{k,\BZ}\frac{1}{8k(k^2+m^2)} = \frac{1}{2 \pi^2} \int_0^\pi 
        \frac{k^2 dk}{8k(k^2+m^2)} + \int_{k,\BZ} \frac{1}{8k(k^2+m^2)} 
        \,\Theta(|k|-\pi) .
\end {equation}
The first integral is easy and gives $\ln(\pi/m)/16 \pi^2$ plus
terms power suppressed in $m$.  When added to
$(-1/16 \pi^2) (\ln(6/m) + 1/2)$,
this cancels the $\ln(m)$, leaving $(1/16\pi^2)(\ln(\pi/6) - 1/2)$.
The final integral has had the small $k$ part of the
integration range removed, so the $m \rightarrow 0$ limit is trivial.
It can then be reduced to
\begin {equation}
\frac{1}{16 \pi^2} \int \frac{d\Omega}{4 \pi} \ln(R({\rm cube})-R({\rm
        ball})) = \frac{1}{16 \pi^2} \frac{12}{\pi} \int_0^{\pi/4}
        d\phi \int_0^{{\rm arctan}( \sec \phi)} \sin(\theta) \,d\theta 
        \ln(\sec(\theta)) 
\end {equation}
which numerically equals $.19233513195/16 \pi^2$.  Note that at no point
have we had to deal numerically with an integral which is log divergent
in $m$, or which still contains $m$ at all.

Combining terms gives $C_3 = -.86147916$, unless we use the unimproved
lattice Laplacian, in which case it is $C_{3,\U} = .08848010$.


\subsection{Three-loop result for \boldmath$\delta\phi^2$}

\begin{figure}[b]
  \centerline{\epsfxsize=10.5cm\epsfbox{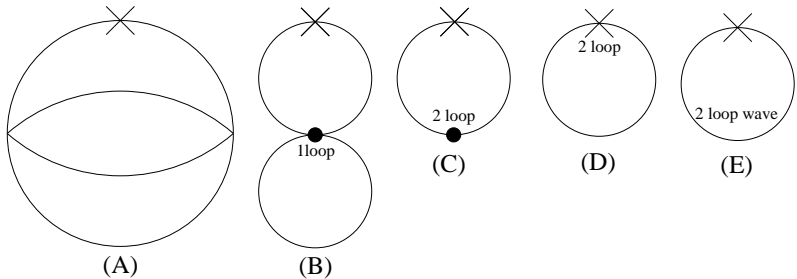}}
  \vspace{0.1in}
  \caption{
    Three-loop vacuum diagrams needed for $\delta \phi^2$
    at 3 loops.  There are 7 additional diagrams either involving 1 
    loop mass counterterms or tadpoles, which cancel among themselves.
    \label{3_loop_graphs}
  }
\end{figure}

In principle, one could extend all the renormalization counterterms
we have considered to three loop order.  However, for many of them,
this requires including mixing of different dimensions of operator
insertions and including counterterms for radiatively induced
high-dimension operators in the Lagrangian.
However, the three-loop contribution to the additive counterterm
$\delta\phi^2$ is an exception: it gives the $O(a)$ corrections
to $\Delta\phiphic$, and we've already seen that we can ignore
higher-dimensional operators at this order.
The calculation of $\delta\phi^2$ is the
least complicated of the 3-loop calculations we might envision, and does 
not require the result of any other 3-loop calculations or any of the
new counterterms which would be needed in a complete 3-loop matching.  
The relevant diagrams are given in Fig.\ \ref{3_loop_graphs}.
We find
\begin{eqnarray}
&&
  \delta\phi^2_{3l}-\delta\phi^2_{2l} =
\nonumber \\ && \qquad
  \left[ \left(N+2\over6\right)^2 \xi^2\Sigma
      + {(N+2)\over 18} \left(
           C_4 - 3 \Sigma C_1 - \Sigma C_2 + \xi \log(a\bar\mu) \right)
        \right] {N u^2 \over (4\pi)^3}
        - \frac{\xi}{4 \pi} \, N r .
\label{eq:dphi3l}
\end{eqnarray}
Here the new constant $C_4$ arises from the basketball diagram of
Figure \ref{3_loop_graphs}, together with part of the mass-squared
counterterm diagram.  The explicit renormalization scale dependence in
(\ref{eq:dphi3l})
cancels the implicit dependence of $r=r(\bar\mu)$ in the last term.

Our definition of $C_4$ is that 
$(C_4 + \xi \log(a\mu))/(4 \pi)^3$ equals the lattice value of the
basketball diagram, (A) in Fig. \ref{3_loop_graphs}, minus diagram (C)
taking only the setting sun part of the two loop mass counterterm.
Explicitly, 
\begin{eqnarray}
\label{C4_int}
\frac{C_4 + \xi \log(a\mu)}{(4 \pi)^3} & = & 
        \int_{k,\BZ} \frac{1}{(\tilde{k}^2)^2} 
        \left( \int_{p,\BZ} \frac{1}{\tilde{p}^2} 
        \Big[ \IL(p{+}k) - \IL(p) \Big] 
        + \int_{p,\Re^3} \frac{1}{p^2} \IC(p) \right)
        \nonumber \\
&& - \int_{k,\Re^3} \frac{1}{k^4} \int_{p,\Re^3} \frac{1}{p^2}
        \, \IC(p{+}k) \, .
\end{eqnarray}
Here $\IL(p{+}k)$ and $\IC(p{+}k)$ arise from the basketball diagram, while
$\IL(p)$ and $\IC(p)$ are from the setting sun piece of the
counterterm.  Every term here is implicitly IR regulated by a common
infinitesimal mass on every propagator, but the way we will perform the
integrations means that we will never need this regulation explicitly.
Also, continuum integrals are implicitly renormalized in $\overline{\rm
MS}$, which will be relevant.

It is convenient to add and subtract the appropriate factor to put the
$\IC(p{+}k)$ factor inside the $\BZ$ $k$ integral.  We add and subtract
\begin{equation}
\left[ \int_{k,\BZ} \frac{1}{(\tilde{k}^2)^2} - 
        \int_{k,\Re^3} \frac{1}{k^4} \right]
        \int_{p,\Re^3}\frac{1}{p^2} \, \IC(p{+}k)
         \, .
\label{sub_off}
\end{equation}
The $p$ integral can be evaluated in $\overline{\rm
MS}$, (note that we do not need its value in the deep IR where the
implicit mass regularization becomes important), and gives
\begin{equation}
\int_{p,\Re^3} \frac{1}{p^2} \IC(p{+}k)
        = \frac{1.5 - \log(k/\mu)}{16 \pi^2} \, .
\end{equation}
The constant terms can be pulled out of the $k$ integrations; the
remaining $k$ integral is Eq.~(\ref{xi_is}) and gives
$\xi / 4 \pi$.  The constant parts
therefore yield $(1.5 + \log(a\mu))\xi/(64 \pi^3)$.  The $\ln k$ leads
to the integral
\begin{equation}
\frac{1}{16 \pi^2} \left( \int_{k,\BZ} \frac{-\ln(ak)}{(\tilde{k}^2)^2} - 
        \int_{k,\Re^3} \frac{-\ln (ak)}{k^4} \right) 
        = \frac{0.30837}{64\pi^3}
\end{equation}
for the improved Laplacian.
Eq.~(\ref{C4_int}), after subtracting Eq.~(\ref{sub_off}), becomes
\begin{equation}
\int_{k,\BZ} \frac{1}{(\tilde{k}^2)^2} \left(
        \int_{p,\BZ} \frac{1}{\tilde{p}^2} \left[\IL(p{+}k)-\IL(p) \right]
        -\int_{p,\Re^3} \frac{1}{p^2} \left[\IC(p{+}k)-\IC(p) \right] 
        \right) \, .
\end{equation}

It is convenient
to split off the part of the continuum 
$p$ integration which lies outside the 
Brillouin zone as a separate integration, which does not suffer from
divergences:
\begin{equation}
 \frac{1}{8} \int_{k,\BZ} \frac{1}{(\tilde{k}^2)^2}
        \int_{p,\Re^3-\BZ} \left( \frac{1}{p^3} - \frac{1}{p^2 |p{+}k|}
        \right) = \frac{.00031757}{64 \pi^3} \, .
\end{equation}
This leaves as the final integral we must consider,
\begin{equation}
\int_{k,\BZ} \frac{1}{(\tilde{k}^2)^2} \int_{p,\BZ} \left(
        \frac{1}{\tilde{p}^2} \left[ \IL(p{+}k)-\IL(p)
        \right] - \frac{1}{p^2} \left[ \IC(p{+}k)-\IC(p)
        \right] \right) \, .
\label {eq:MonteMe}
\end{equation}
This integration is finite and over a finite integration region
(provided we make a prescription that the argument of the $p$ integral
is interpreted as the average over $p$ and $-p$ of the quantity written).
The integration is nine dimensional and contains some delicate integrable
singularities.  We find it convenient to use adaptive Monte Carlo
integration rather than quadratures.  Monte Carlo integration is
somewhat problematical when there are integrable singularities.
Rather than finding clever changes of variables to get rid of such
singularities, the simplest thing to do is to include a small
mass in all the propagators (thus cutting off all singularities),
vary that mass, and then numerically extrapolate the zero mass limit
from the results of the Monte Carlo integrations.  Our result for
the integral (\ref{eq:MonteMe}) is
$(0.0985\pm0.006)/(4\pi)^3$.  Combining terms,
we then have $C_4 = 0.2817(6)$ for the improved Laplacian.


\subsection {Minimalist expression for \boldmath$\Delta\phiphic$ through
             \boldmath$O(a)$}

Some of the development in the preceding sections is unnecessary for
the isolated goal of getting an $O(a)$ improvement of
$\Delta\langle\phi^2\rangle$.
It is possible to combine the previous results in the more compact form
\begin {mathletters}%
\label{eq:match}%
\begin {equation}
   \Delta\langle\phi^2\rangle_\cont
   =
   {Z_r\over Z_\phi} \, \Delta\langle\phiz^2\rangle_\lat
        + {N \xi\over4\pi} \, a r_1
        - {N(N+2)\over18} \, {{\cal C}\over (4\pi)^3} \, a u_0^2
   + O(u^3) ,
\end {equation}
\begin {equation}
   {\cal C} \equiv C_4 + \xi \ln 6 + \xi C_3 .
\end {equation}
Here $\phiz$ is the {\it bare}\/ lattice field, and $u_0$ the bare
lattice coupling, corresponding to
the bare lattice action (\ref{eq:S0});
\begin {equation}
   \Delta\langle\phiz^2\rangle_\lat
         = \langle\phiz^2\rangle_\lat - {N \Sigma\over 4\pi a} \,;
\end {equation}
$r_1$ stands for the tadpole-adjusted mass,
\begin {equation}
   r_1 \equiv r_0 + {(N+2)\over 6}  \, u_0 \, {\Sigma\over4\pi a} ,
\end {equation}
which is $O(u^2)$ near the transition;
and $r_0$ is the bare mass used in the bare lattice action (\ref{eq:S0}).
We have retreated from lattice units and explicitly show all
factors of $a$.
Since $\Delta\langle\phiz^2\rangle$ is $O(u)$ near the transition,
we only need the one-loop result
\begin {equation}
   {Z_r\over Z_\phi} \simeq 1 + {(N+2)\over6} \, u a \, {\xi\over 4\pi}
\label {eq:Zratio}
\end {equation}
for $Z_r/Z_\phi$.

To finish the relationship between the measurement of
$\Delta\langle\phi^2\rangle/u$ and the bare fields and parameters of
the lattice Lagrangian, we only need the one-loop relationship between
the continuum and bare lattice couplings:
\begin {equation}
   u_0 \simeq u  + {(N+8)\over 6} \, u^2 a \, {\xi\over4\pi}.
\end {equation}
\end {mathletters}

The form (\ref{eq:match}) has the conceptual advantage of not introducing the
renormalization scale $\bar\mu$, since its introduction is unnecessary if
one's
only interest is
in $\Delta\phiphic$ and not the value of $r_c$.
Eq.\ (\ref{eq:match}) also
makes clear that nothing depends on $C_1$ and $C_2$ at this order, so that one
doesn't really need those integrals.  And it makes clearer that
there is no $O(a^0)$ correction to $\Delta\langle\phi^2\rangle$, which
is obscured by eqs.\ (\ref{eq:phi2renorm}) and (\ref{eq:dphimaster}).
That is,
\begin {equation}
   \Delta\langle\phi^2\rangle_\cont
   =
   \Delta\langle\phiz^2\rangle_\lat
   + O(a u^2) ,
\end {equation}
as in (\ref{eq:phi2renorm0}).

However, this is not how we have implemented our calculation of
$\Delta\phiphic$ when quoting numbers from simulations.  What we have
done is described in the text and the earlier parts of this
appendix and, though equivalent through $O(a)$, will
give slightly different numerical values because of differences in
higher orders in $a$.


\section {Logs in Large Volume Scaling for \boldmath$\alpha=0$}
\label {app:log}

In this section, we review standard renormalization group arguments about
the free energy and verify that a $t^2 \ln L$ term appears in the free
energy when $\alpha = 0$.
If one increases renormalization scale by a ``blocking'' factor of $b$,
the free energy density
of the blocked system is related to the free energy of the
original system by a transformation law of the form
\begin {equation}
   {\cal F}(\{K\}) = {\cal G}(\{K\}) + b^{-d} {\cal F}(\{K'\}) ,
\end {equation}
where ${\cal F}$ is the free energy per block,
$\{K\}$ represents the couplings of the theory,
and ${\cal G}$ is an analytic function (depending on $b$) that
represents the contributions to the free energy from the degrees of
freedom that have been blocked.%
\footnote{
   Our presentation follows, for example, chapter 3 of Ref.\ \cite{cardy},
   which is one of many
   nice introductions to the renormalization group.  Our ${\cal F}$ and
   ${\cal G}$ are that reference's $f$ and $g$.
   We use ${\cal F}$ to
   avoid confusion with $f$ in the text,
   which was the free energy per physical volume rather than per block.
}
Iterating $n$ times,
this becomes
\begin {equation}
   {\cal F}(\{K\}) = \sum_{j=0}^{n-1} b^{-jd} {\cal G}(\{K^{(j)}\})
                  + b^{-nd} {\cal F}(\{K^{(n)}\}) ,
\label {eq:iterate}
\end {equation}
where $\{ K^{(j)} \}$ is the $j$th iterate of $\{K\}$.
If we start with a block size of $a$ and iterate all the way out to
size $L$, we have $n = \log_b(L/a)$.

For $t$ very small but non-zero, the description of the system will first flow
towards the critical point, as one blocks to larger and larger distances.
But it will then eventually flow away from the fixed point,
closely following one of the two
unique ``outflow'' trajectories from the fixed point
(one for $t>0$ and one for $t<0$), corresponding to all irrelevant couplings
being set to zero.
The inhomogeneous part ${\cal G}$ of the
transformation law (\ref{eq:iterate}) will generate singular behavior as
$t \to 0$, which may then be written as
\begin {equation}
   {\cal F}(t) \sim \sum_{j=0}^{\log_b (L/a)} b^{-jd} {\cal G}(b^{j y_t} t) ,
\end {equation}
where $b^{j y_t} t$ parameterizes flow along the outflow trajectory.
For the infrared behavior at large $L$ and small $t$, we can replace the sum
by an integration.  Changing integration variable to $s \equiv b^{j y_t} t$,
\begin {equation}
   {\cal F}(t) \sim {t^{d/y_t} \over y_t \ln b} \int_t^{t (L/a)^{y_t}}
                 s^{\alpha-3} {\cal G}(s) \, ds ,
\end {equation}
where $\alpha = 2 - d/y_t$.  For $\alpha=0$, this is
\begin {equation}
   {\cal F}(t) \sim {2 t^2 \over d \ln b} \int_t^{t (L/a)^{d/2}}
                 s^{-3} {\cal G}(s) \, ds .
\end {equation}
To help tame the singularity as $s \to 0$, integrate by parts two times.
This leaves an integral proportional to
\begin {equation}
   t^2 \int_t^{t (L/a)^{d/2}} s^{-1} {\cal G}''(s) \, ds
\end {equation}
plus terms which fit the too-naive scaling form (\ref{eq:fnaive}).
The remaining integral can be rewritten as
\begin {equation}
   {\cal G}''(0) \, t^2 \ln(L^{-d/2})
   + t^2 \int_t^{t (L/a)^{d/2}}
                 s^{-1} \left[{\cal G}''(s) - {\cal G}''(0)\right]\, ds .
\end {equation}
The second term has the desired analyticity properties of the
too-naive scaling form (\ref{eq:fnaive}), and the first term is the
logarithm of Privman and Rudnick appearing in (\ref{eq:privman}).


\section {Our own analysis of $C_\c$}
\label {app:Cc}

\subsection {Numerical simulation}

In this appendix, we discuss our own attempt to determine $C_\c$, to make
a crude check of the value we take from Ref.\ \cite{campostrini}.
The scaling of the intersection values
$C_\times(L_1,L_2)$ of cumulant curves $C(r)$ is
\begin {equation}
   C_\times(L_1,L_2) - C_\c
   \sim { L_1^{y_t} L_2^{-\omega} - L_2^{y_t} L_1^{-\omega} \over
          L_1^{y_t} - L_2^{y_t} }
   =  {b^{-\omega} - b^{y_t} \over 1 - b^{y_t}} \, L_1^{-\omega}
\label {eq:Cscaling}
\end {equation}
for $L_1,L_2 \to \infty$.
This is a simple consequence of Binder's analysis
\cite{BinderCumulants},
but, for the sake of completeness, we briefly outline the argument in
Appendix \ref{app:cumulant}.

We would like to choose $L$ as large as
possible, in order to make the scaling law (\ref{eq:Cscaling}) as
accurate as possible, so that we can use it to extrapolate a good
value of $C_\c$.
Our ultimate interest in this paper is to study continuum O(2) theory,
which requires $ua \ll 1$ and for which the scaling limit is
$L/a \gg 1/ua$.  But $L/a \gg 1/ua$ implies that small $ua$ simulations
are an inefficient choice
for getting as far as possible into the scaling limit.
Because $C_\c$ is universal, we can extract it from large $ua$ simulations
rather than small $ua$ simulations.
[In fact we could use any model in the same universality class.]
Our simulation code is optimized to perform best if
$ua$ is not extremely large, and so we have chosen to extract $C_\c$ from
data taken with $u_0a=60$.
Because $u_0a=60$ is not a small value of $ua$, the $O(a)$ improvements
to the action are pointless; unlike other simulations reported in
this paper, we quote quantities in terms of the bare lattice parameters.

\begin {figure}
\vbox{
   \begin {center}
      \epsfig{file=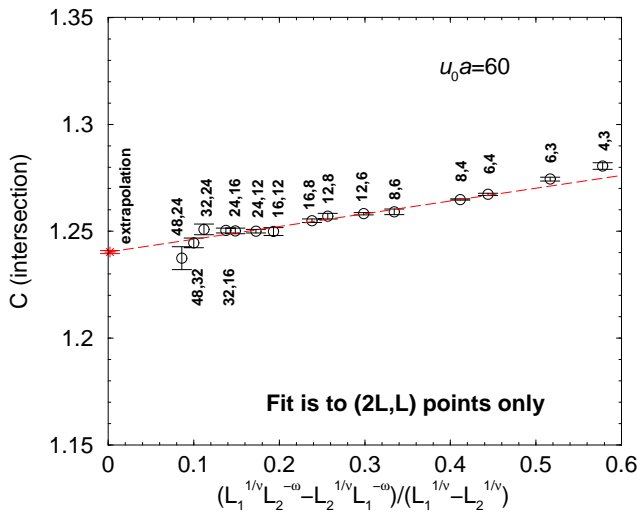,scale=.40,angle=-90}
   \end {center}
   \caption{
       The values $C_\times(L_1,L_2)$ corresponding to the intersection
       points of pairs of the curves $C(r)$ for different lattice sizes.
       The horizontal axis has been chosen so that this relationship should
       be linear as $L_1, L_2 \to \infty$ (which corresponds
       to approaching zero on the horizontal axis).  Each point is labeled
       by $L_1/a$ and $L_2/a$.  The linear fit shown is only to that subset
       of points with $L_1 = 2 L_2$.
       \label{fig:Cscaling}
   }
}
\end {figure}

A variety of
intersection values $C_\times(L_1,L_2)$ of $C(r)$ curves are plotted in
Fig.\ \ref{fig:Cscaling} against
$( L_1^{y_t} L_2^{-\omega} - L_2^{y_t} L_1^{-\omega} )/
         ( L_1^{y_t} - L_2^{y_t} )$
which, by (\ref{eq:Cscaling}), should lead to a linear relationship
at large $L_1,L_2$.  The errors on points sharing an $L$ value
are correlated, and we compute and use the full correlation
matrix for making fits.  For the sake of simplicity, however, we have
only fit the subset of data with $L_1 = 2 L_2$, with results for
$C_\c$ and the associated confidence levels given in Table
\ref{tab:Cfits}.  If we naively include smaller and smaller $L$ until
the confidence level of the fit becomes poor, we would fit all the way
down to $L=4$ and obtain
\begin {equation}
   C_\c = 1.2402(7) 
\label{eq:Ccrit}
\end {equation}
as our final result.
We are suspicious that this result may have a ``tail wagging the dog''
systematic error.  Our small $L$ data has smaller statistical error than our
large $L$ data and may be over-weighted in the fit to what's supposed to
be large $L$ asymptotic behavior.
(In particular, in an earlier analysis where we had even poorer statistics
on the large $L$ data, we found a result that disagreed even more drastically
from the result of Ref.\ \cite{campostrini}.)

\begin {table}
  \begin {center}
  \setlength{\tabcolsep}{10pt}
  \begin {tabular}{|l|l|l|}                          \hline
  \multicolumn{1}{|c|}{fit} &  \multicolumn{1}{c|}{$C_\c$}   & C.L. \\ \hline
     $L_2/a \ge 16$ & 1.229(9)  & \multicolumn{1}{c|}{NA}  \\
     $L_2/a \ge 12$ & 1.243(4)  &  6\% \\
     $L_2/a \ge 8$ & 1.2404(22) & 13\% \\
     $L_2/a \ge 6$ & 1.2410(15)&  25\% \\ 
     $L_2/a \ge 4$ & 1.2402(7) &  35\% \\
     $L_2/a \ge 3$ & 1.2394(7)  & 1\% \\
            \hline
  \end {tabular}
  \end {center}
  \caption{
     Results of linear fits to the subset of data of fig.\
     \protect\ref{fig:Cscaling}
     with $L_1 = 2L_2$, and the corresponding confidence levels
     (``NA'' means not applicable).
     The first row corresponds to a fit to two points, the second to
     three points, etc.
     \label{tab:Cfits}
  }
\end {table}


\subsection {Large volume scaling of cumulant intersections}
\label {app:cumulant}

Let's quickly reproduce Binder's result (\ref{eq:rscaling})
for the scaling of $r_\times(L_1,L_2)-r_\c$.
For large $L$, we will focus on the scaling piece of the cumulant which, in the
notation of section \ref{sec:Lscaling}, is
\begin {equation}
   C(t,\{u_j\},L^{-1})
      \sim C_\sing(b^{y_t} t, \{u_j b^{y_j}\}, b/L) .
\end {equation}
Choosing $b = L$,
\begin {equation}
   C \simeq C_\sing(L^{y_t} t, \{u_j L^{y_j}\}, 1) ,
\end {equation}
where $C_\sing$ is a universal scaling function.
Now treat $L^{y_t} t$ and $u_j L^{y_j}$ as small and Taylor expand,
keeping track of only the most important irrelevant operator $u_j$,
\begin {equation}
   C \simeq C_\sing(0,0,1) + A L^{y_t} t
      + B L^{-\omega} .
\end {equation}
Note that $A$ and $C_\sing(0,0,1)$ are universal, but $B$ is not, because
it depends on the value of $u_j$.
Now look for the intersection for two different system sizes:
\begin {equation}
   C_\sing(0,0,1) + A L_1^{y_t} t + B L_1^{-\omega}
   =
   C_\sing(0,0,1) + A L_2^{y_t} t + B L_2^{-\omega} ,
\label {eq:Cexpand}
\end {equation}
which has solution
\begin {equation}
   t \simeq {B(L_1^{-\omega}-L_2^{-\omega}) \over
        A(L_2^{y_t}-L_1^{y_t})}
\label {eq:texpand}
\end {equation}
up to corrections suppressed by additional powers of $1/L$.
This is just the scaling (\ref{eq:rscaling}) quoted for $r_\times-r_\c$.
(And one sees {\it a posteriori}\/ that it was justified to treat $L^{y_t}t$
as small in the $L \to \infty$ limit.) 
All that is necessary to derive the scaling (\ref{eq:Cscaling}) of
$C_\times$ is to plug (\ref{eq:texpand}) back into (\ref{eq:Cexpand})
and rename $C_\sing(0,0,1)$ as $C_\c$.

In the main text, our procedure for defining the nominal transition in
finite volume, for our small $ua$ simulations, is to find the point
where $C = C_\c$.  From (\ref{eq:Cexpand}), one may verify that this
prescription gives
\begin {equation}
   t \simeq {B\over A} \, L^{-{y_t}-\omega} \sim L^{-{y_t}-\omega} ,
\end {equation}
as asserted in the main text.


\section {Small \boldmath$L$\lowercase{\boldmath$u$} expansion
          of \boldmath$\Delta\phiphic$}
\label {app:smallLu}

In this appendix, we compute the expected result for $\Delta\phiphic$
for small volumes in the continuum limit, where we define the nominal
critical temperature in such volumes using Binder cumulants, as explained
in Sec.\ \ref{sec:binder}.
We have repeatedly described
``short distance'' physics (distance small compared to $1/u$) as
perturbative, and so for small volumes ($L \ll 1/u$), one might first try
simple perturbation theory to compute $\Delta\langle\phi^2\rangle$.
The leading-order diagram contributing to $\Delta\langle\phi^2\rangle$
is shown in Fig.\ \ref{one_loop_graphs}d
and corresponds to
\begin {equation}
   \langle\phi^2\rangle \simeq {N\over L^3} \sum_\p {1\over p^2 + r} ,
\label {eq:naivepert}
\end {equation}
where for the sake of generality we have considered an O($N$) model,
and the actual case of interest is $N=2$.  The sum is over the discrete
momenta $\p = 2\pi \n/L$ associated with the $L\times L\times L$ periodic
volume.
There is a problem with (\ref{eq:naivepert}), however.  Consider the
mean-field theory approximation $r_\c = 0$ to the critical value of $r$.
The sum then has a infrared divergent term associated with $\p = 0$.
The problem
is that it is more accurately large momentum physics which is perturbative,
rather than small distance physics.  So, even when $L \to 0$, the
physics associated with the $\p = 0$ modes is non-perturbative.
Since this is only one mode, we simply analyze it separately, and then treat
all the $\p \not= 0$ modes perturbatively.


\subsection {The case \boldmath$r=0$: leading order}

   To illustrate the calculation, let us first compute
$\Delta\langle\phi^2\rangle$ if $r = 0$.  We'll later come back to
consider the actual $r$ that is chosen by the criteria of our simulations
that the cumulant $C = C_\c$.  Consider the approximation $S_0$ to the
action where we ignore everything but the $\p{=}0$ modes $\phi_0$:
\begin {equation}
   S_0
   = \int d^3 x \> \left[ \frac 12 |\grad\phi_0|^2 
	+ {u\over 4!} |\phi_0|^4 \right]
   = {L^3 u\over 4!} \, |\phi_0|^4 .
\end {equation}
Note that this result holds on the lattice as well as in the continuum.
The $\p = 0$ contribution to $\Delta\langle\phi^2\rangle$ is then
\begin {equation}
   \langle\phi_0^2\rangle
   \simeq
     {\int d^N\phi_0 \> e^{-S_0} |\phi_0|^2 \over
      \int d^N\phi_0 \> e^{-S_0} }
   = \left(4! \over L^3 u\right)^{1/2} \,
      {\int d^N x\> e^{-x^4} x^2 \over \int d^N x\> e^{-x^4}}
   = \left(4! \over L^3 u\right)^{1/2}
      {\Gamma\left(N+2\over 4\right) \over \Gamma\left(N\over 4\right)} \,.
\label {eq:zmode0}
\end {equation}
By dimensional analysis, the leading perturbative contribution of the
$\p\not=0$ modes to $\langle\phi^2\rangle$ should be order $L^{-1}$ and
so is dominated by the zero-mode contribution (\ref{eq:zmode0}).
Specializing to $N=2$, we then have
\begin {equation}
   {\langle\phi^2\rangle_{r=0} \over u}
   = {\sqrt{4!/\pi} \over (Lu)^{3/2}} + O\blparen(Lu)^{-1}\brparen
   \simeq {2.76395 \over (Lu)^{3/2}} + O\blparen(Lu)^{-1}\brparen .
\label {eq:smallL0}
\end {equation}
As we'll see explicitly below, the UV subtraction that converts
$\langle\phi^2\rangle$ to $\Delta\langle\phi^2\rangle$ in the
continuum limit is not
relevant until next order in $(Lu)^{-1}$, and so
$\Delta\langle\phi^2\rangle/u$ also has the expansion (\ref{eq:smallL0}).


\subsection {The case \boldmath$r=0$: next-to-leading order}

Now consider the leading-order perturbative contribution
$\delta\langle\phi^2\rangle$ of the
non-zero modes to $\langle\phi^2\rangle$ from (\ref{eq:naivepert}):
\begin {equation}
   \delta\langle\phi^2\rangle
   \simeq {N \over L^3} \sum_{\p\not=0} {1\over p^2} \,.
\label {eq:pert1}
\end {equation}
This sum is UV divergent, and the UV divergence is simply the free-field
divergence discussed in section (\ref{sec:div0}).  It is subtracted when
we compute $\Delta\langle\phi^2\rangle$ as opposed to $\langle\phi^2\rangle$.
Note that our prescriptions (\ref{eq:phi2renorm0}) or (\ref{eq:phi2renorm})
for computing $\Delta\langle\phi^2\rangle$ in our simulations
involve subtracting the
{\it infinite volume}\ free-field result for $\langle\phi^2\rangle$.
In the continuum limit, this subtraction turns the perturbation
(\ref{eq:pert1}) into
\begin {equation}
   \delta\Delta\langle\phi^2\rangle =
   {N\over L^3} \sum_{\p\not=0} {1\over p^2}
       - N \int_\p {1\over p^2}
   = {N\over (2\pi)^2 L} \left[
        \sum_{\n\not=0} {1\over n^2} - \int {d^3n\over n^2}
     \right] .
\end {equation}
Individually, the sums and integrals above must be consistently regulated,
for example by dimensional regularization or by keeping the system
on a lattice with arbitrarily small lattice spacing.

There are a number of ways to evaluate the result numerically.
One is to start by regulating with dimensional regularization, working in $d$
spatial dimensions.  Then
\begin {equation}
   \delta\Delta\langle\phi^2\rangle =
   {N\over L^d} \sum_{\p\not=0} {1\over p^2}
       - N \int_\p {1\over p^2}
   = {N\over (2\pi)^2 L^{d-2}} \left[
        \sum_{\n\not=0} {1\over n^2} - \int {d^dn\over n^2}
     \right] .
\end {equation}
Now rewrite
\begin {equation}
   {1\over n^2} = \int_0^\infty ds \, e^{-s n^2} .
\end {equation}
The $n_1$, $n_2$, ... sums and integrals then factor, giving
\begin {equation}
   \sum_{\n\not=0} {1\over n^2} - \int {d^dn\over n^2}
   = \int_0^\infty ds\> \left\{ \left[ \theta_3(e^{-s}) \right]^d - 1
            - \left(\pi\over s\right)^{d/2}\right\} ,
\label{eq:ellipticd}
\end {equation}
where
\begin {equation}
   \theta_3(q) \equiv \sum_{k=-\infty}^\infty q^{(k^2)}
\end {equation}
is a special case of an elliptic theta function.  The integral on
the right-hand side of (\ref{eq:ellipticd}) is absolutely convergent
in $d=3$, so we can now dispense with dimensional regularization and
set $d$ to three.  The integral is then easily done numerically, giving
\begin {equation}
   \sum_{\n\not=0} {1\over n^2} - \int {d^dn\over n^2}
   \simeq -8.91363 .
\end {equation}
For $N=2$, one then has
\begin {equation}
   {\delta\Delta\langle\phi^2\rangle_{r=0} \over u}
   \simeq - {0.451570\over Lu} .
\label{eq:del}
\end {equation}

The effect of interactions in the non-zero mode sector will be perturbative
for small $Lu$ and will give higher order contributions to
$\Delta\langle\phi^2\rangle$.  However, we must also consider
interactions of the zero mode $\phi_0$ with the non-zero modes, which will
give corrections to the action $S_0$ used in our earlier analysis of
$\phi_0$.  For instance, the coupling $u$ in $S_0$ will pick up corrections
of order $L u^2$.  This will generate an $O(Lu)$ relative
correction to the zero-mode
contribution (\ref{eq:smallL0}),
and so that correction will be higher order than the
contribution (\ref{eq:del}) computed
above.  A similar story will hold for corrections to $r$ (and for the effects
of higher-dimensional interactions) with the complication that $r$ will
receive some infinite contributions in the continuum limit, corresponding to
the usual mass renormalization.  The latter is absorbed by the usual
renormalization of $r$, and so one has
\begin {equation}
   {\delta\langle\phi^2\rangle_{r=0} \over u}
   \simeq {2.76395 \over (Lu)^{3/2}}
      - {0.451570\over Lu} + O\blparen(Lu)^{-1/2}\brparen
\end {equation}
if one interprets the $r$ in the condition $r=0$ as being renormalized $r$.
(The details of the renormalization scheme will not matter at the order shown.)


\subsection {The case \boldmath$C=C_\c$}

Now, instead of setting $r = 0$, we will choose $r$ so that the cumulant
$C$ is equal to its critical value.  Let's begin with a leading-order analysis
and so focus on just the $\p=0$ sector.  For general $r$, we then have
\begin {equation}
   S_0
   = {L^3 r \over 2} \, |\phi_0|^2 + {L^3 u\over 4!} \, |\phi_0|^4
\end {equation}
and
\begin {equation}
   \langle\phi_0^{2k}\rangle
   = {\int d^N\phi_0 \> e^{-S_0} |\phi_0|^{2k} \over
      \int d^N\phi_0 \> e^{-S_0}} .
\end {equation}
By a change of variables, this can be rewritten as
\begin {equation}
   \langle\phi_0^{2k}\rangle = \left(4!\over L^3 u\right)^{k/2}
        {I_k(R) \over I_0(R)}
\end {equation}
where
\begin {equation}
   I_k(R) \equiv \int_0^\infty dy \> e^{-y^2-R y} y^{k+(N-2)/2} ,
\end {equation}
and
\begin {equation}
   R \equiv {r\over 2} \left( 4! L^3 \over u\right)^{1/2} .
\label {eq:Rdef}
\end {equation}

Now note that the volume average $\bar\phi$ of the field, used in the
definition of the cumulant, is simply $\phi_0$.
Specializing now to $N=2$ and doing the $I_k$ integrals, one obtains
\begin {eqnarray}
   I_0(R) &=& {\sqrt\pi\over2} \, e^{R^2/4} {\rm erfc}\!\left(R\over2\right) ,
\\
   I_1(R) &=& \half [1 - R\, I_0(R)] ,
\\
   I_2(R) &=& \fourth [-R + (R^2+2) I_0(R)] .
\end {eqnarray}
The cumulant can be written
\begin {equation}
   C \equiv {\langle\phi_0^4\rangle \over \langle\phi_0^2\rangle^2}
     \simeq {I_2(R) \, I_0(R) \over [I_1(R)]^2} \,.
\end {equation}
A numerical search for the point where $C=C_\c\simeq1.243$
gives
\begin {equation}
   R \simeq -2.5073 \, .
\end {equation}
At this $R$, the zero-mode contribution to $\langle\phi^2\rangle$ is then
\begin {equation}
   \langle\phi^2\rangle_\c
    = \left(4!\over L^3 u\right)^{1/2} {I_1(R) \over I_0(R)}
           + O\blparen(Lu)^{-1}\brparen
    = {6.44003 \over (L^3 u)^{1/2}} 
           + O\blparen(Lu)^{-1}\brparen .
\label{eq:smallLuC0}
\end {equation}

Since the result for $R$ is a pure number, the definition (\ref{eq:Rdef})
of $R$ shows that $r$ is of order $\sqrt{u/L^3}$, which is smaller
by a power of $\sqrt{Lu}$ than any non-zero momentum squared, which are
order $L^{-2}$.  So $r$ can be ignored in finding the leading contribution
of the non-zero modes, with the effect that $\delta\Delta\langle\phi^2\rangle$
is the same as in the earlier $r=0$ analysis. 
Adding (\ref{eq:smallLuC0}) and (\ref{eq:del}) then produces the result
(\ref{eq:smallLu}) quoted in the main text.


\section {System size and the simulations of Gr\"uter \etal}
\label {app:gruter}

One of the applications of O(2) field theory is to studying the corrections
to the critical temperature, due to interactions, for Bose-Einstein
condensation of a nearly-ideal non-relativistic Bose gas.
This application has been previously studied using numerical techniques.
Our discussion of system size in section \ref{sec:Lbig} has
implications for an early study by Gr\"uter \etal \cite{gruter},
which is that much of the data collected was likely in
insufficiently large volume.
These simulations did not make use of O(2) field theory:
They worked in the canonical ensemble and studied the path integral
for a fixed number $\Np$ of particles in a finite volume $V$.
If no attempt were made to fit for large volume corrections, then
Fig.\ \ref{fig:ua6} makes it clear that one should take
roughly $Lu \ge 400$ to keep those corrections moderately small.
It's illuminating to also consider the less restrictive condition of roughly
$Lu \ge 200$.
We can translate these conditions on system size to the context of
Bose-Einstein condensation by translating the parameter $u$ of the
O(2) field theory.
For this application, the relation is \cite{baym1}
$u = 96 \pi^2 a/\lambda^2$, where $a$ is the scattering length of the
atoms, $\lambda = \hbar \sqrt{2\pi/m\kB T}$ is the thermal wavelength, and $m$
is the mass of the atoms.
Using the ideal gas approximation
$T_\c \simeq T_0 = (2\pi\hbar^2/\kB m) [n/\zeta\!\left(3\over2\right)]^{2/3}$
for the critical temperature,
one can write
\begin {equation}
   n a^3 \simeq \left(Lu\over 96\pi^2\right)^3 {\zeta(\threehalf)^2 \over \Np}
\end {equation}
at the transition.
The conditions $L u \ge 400$ or 200 may then be
translated into the conditions
\begin {equation}
   n a^3 \ge {0.51\over \Np}
   \quad
   \mbox{or}
   \quad
   {0.064\over \Np} \,.
\end {equation}
The largest $\Np$ used in the simulations of Gr\"uter \etal\ was
$\Np=216$, and their extraction of the critical temperature depended
on results with $\Np = 125$ as well.
For the latter, our rough conditions for being
in large enough volume then requires $n a^3 \ge 0.004$ or $0.0005$,
depending on whether one takes the more or less restrictive condition
$Lu \ge 400$ or $Lu \ge 200$.
Gr\"uter \etal\ quote results for $na^3$ all the way down to $10^{-5}$,
and the majority of points in their small $na^3$ tail have $n a^3 \le 10^{-4}$.
(See Fig.\ 3 of Ref.\ \cite{gruter}.)
It therefore seems at least possible that they may have had inadequate
system sizes for their extrapolation of the $ua \to 0$ behavior.


\end {document}